\documentclass[10pt,a4paper]{article} % new guy 
\usepackage{anysize}
\usepackage[format = hang]{caption}
\usepackage{amsmath,amsthm,verbatim,amssymb,amsfonts,amscd, graphicx}
\usepackage{graphics,natbib}
\usepackage{titlesec,footmisc}%,sectsty
\usepackage{hyperref} %url

\usepackage{listings}
\usepackage{booktabs} % for tables in small font? 

\usepackage{color}

\theoremstyle{plain}

\theoremstyle{definition}

\newcommand{\var}{\text{Var}}

\newcommand{\corr}{\text{corr}}

\def\E{{\rm E}}

\def\var{{\rm var}}
\def\bsq{{\rm bsq}}
\def\tr{{\rm t}}
\def\midd{\,|\,}

\def\hatt{\widehat}
\def\arr{\rightarrow}
\def\N{{\rm N}}
\def\half{\hbox{$1\over 2$}}
\def\eps{\varepsilon}

\def\beq{\begin{eqnarray}}
\def\eeq{\end{eqnarray}}

\def\beqn{\begin{eqnarray*}}  % no equation numbers are generated
\def\eeqn{\end{eqnarray*}}

\def\E{{\rm E}}

\def\N{{\rm N}}

\def\Pr{P}

\def\risk{{\rm risk}}
\def\fic{{\rm FIC}}
\def\medfic{{\fic}^m}

\def\rootfic{{\rm rootFIC}}

\def\diag{{\rm diag}}

\def\quadandquad{\quad {\rm and} \quad}
\def\arr{\rightarrow}
\def\hatt{\widehat}
\def\tilda{\widetilde}
\def\sumin{\sum_{i=1}^n}

\def\eps{\varepsilon}
\def\half{\hbox{$1\over2$}}

\def\rootn{\sqrt{n}}

\def\obs{{\rm obs}}
\def\midd{\,|\,}
\def\tr{{\rm t}}
\def\dell{\partial}

\def\true{{\rm true}}

\def\aic{{\rm aic}}

\def\Tr{{\rm Tr}}

\def\mse{{\rm mse}}
\def\rmse{{\rm rmse}}

\def\pois{{\rm Pois}}

\def\narr{{\rm narr}}
\def\wide{{\rm wide}}
\def\dellone{\hbox{$\dell\mu\over \dell\theta$}}
\def\delltwo{\hbox{$\dell\mu\over \dell\gamma$}}

\def\final{{\rm final}}

\titleformat{\section}{\normalfont\large\sc\centering}{\thesection}{1em}{}
\titleformat{\subsection}[runin]{\normalfont\large\bfseries}{\thesubsection}{1em}{}
\numberwithin{equation}{section} 
\renewenvironment{abstract}
   {\list{}{\rightmargin\leftmargin} %
    \item[\text{\hspace{10mm}\sc Abstract.}]\relax}
   {\endlist}
\begin{document}

%% saa og si ferdig 17-xii-2019 
% \def\today{Version 0.99, as of 14-iv-2019}
% \def\heute{Version 0.93, as of 16-xii-2019}
\def\heute{December 2019}

\begingroup
\begin{centering} 
%% 
% \large{\bf CD-FIC and median-FIC: \\ 
% Confidence Distributions for the FIC scores}\\[0.8em]
\large{\bf Confidence Distributions for FIC scores} \\[0.8em]
\large{\bf C\'eline Cunen and Nils Lid Hjort} \\[0.3em] % \par
%% \small {\rm $^1$nils@math.uio.no} \\[0.3em] 
\small {\sc Department of Mathematics, University of Oslo} \\[0.3em]
% \small {\sc P.B.~1053, Blindern, N-0316 Oslo, Norway} \\[0.8em]
\small {\sc {\heute}}\par
\end{centering}
\endgroup

% \keywords{improving on MLE, inadmissibility, loss functions, 
% Poisson parameters, simultaneous estimation}

\begin{abstract}
\small{
When using the Focused Information Criterion (FIC) for 
assessing and ranking candidate models with respect to 
how well they do for a given estimation task, it is 
customary to produce a so-called FIC plot. This plot
has the different point estimates along the y-axis and 
the root-FIC scores on the x-axis, these being the 
estimated root-mean-square scores. In this paper we 
address the estimation uncertainty involved in each 
of the points of such a FIC plot. This needs careful
assessment of each of the estimators from the 
candidate models, taking also modelling bias into 
account, along with the relative precision 
of the associated estimated mean squared error quantities.
We use confidence distributions for these endeavours. 
This leads to fruitful CD-FIC plots,
helping the statistician to judge to what extent the 
seemingly best models really are better than other
models, etc. These efforts also lead to 
two further developments. The first is a new tool for 
model selection, which we call the quantile-FIC, 
which helps overcome certain difficulties associated 
with the usual FIC procedures, related to somewhat 
arbitrary schemes for handling estimated squared biases. 
A particular case is the median-FIC. The second development 
is to form model averaged estimators with fruitful weights 
determined by the relative sizes of the 
median- and quantile-FIC scores. 

}
\end{abstract}

% \keywords{admissibility; Bayes and empirical Bayes; 
%    minimax; Poisson; shrinkage; simultaneous estimation}

%% \def\date{27-Sep-2017}

\section{Introduction and summary}
\label{section:intro}

\begin{figure}[h]
\centering
\includegraphics[scale=0.4]{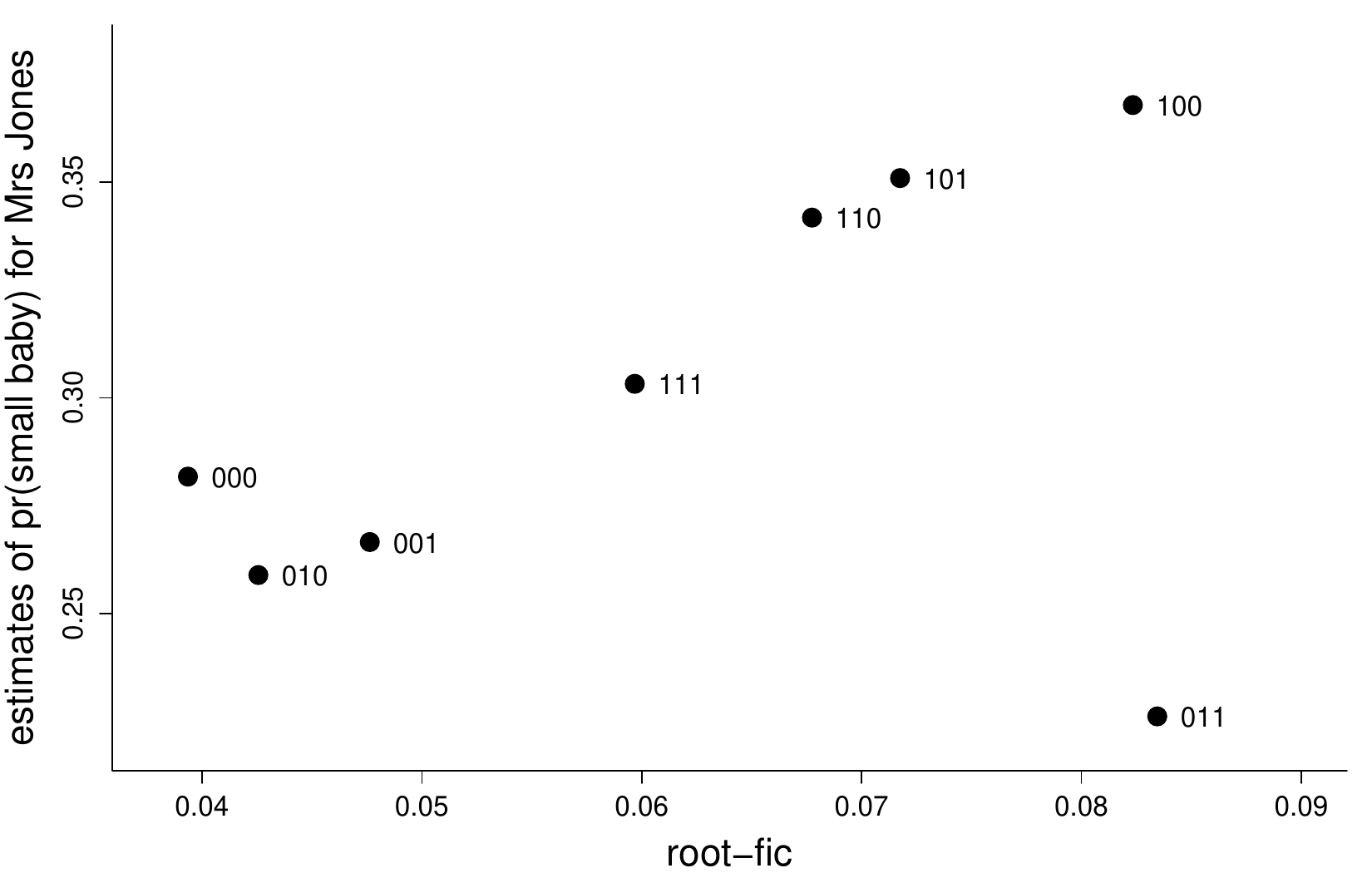}
\caption{FIC plot for the $2^3=8$ models for estimating 
the probability of having a small child, for Mrs.~Jones
(white, age 25, 60 kg, smoker). Here `101' is the model
where $z_1,z_3$ in in and $z_2$ is out, etc.} 
\label{figure:fig14}
\end{figure}

Mrs.~Jones is pregnant. She's white,  
%% [xx of european decent xx]
25 years old, a smoker, and of median weight 60 kg before pregnancy. 
What's the chance that her baby-to-come will be small, 
with birthweight less than 2.50 kg 
(which would mean a case of neonatal medical worry)? 
Figure \ref{figure:fig14} gives a {\it FIC plot}, 
using the Focused Information Criterion to display
and rank in this case $2^3=8$ estimates of 
this probability, computed via eight logistic
regression models, inside the class  
\beqn
%% p(x_1,x_2,z_1,z_2,z_3)
p=\Pr\{y=1\midd x_1,x_2,z_1,z_2,z_3\}
   ={\exp(\beta_0+\beta_1x_1+\beta_2x_2+\gamma_1z_1+\gamma_2z_2+\gamma_3z_3)
   \over 
  1+\exp(\beta_0+\beta_1x_1+\beta_2x_2+\gamma_1z_1+\gamma_2z_2+\gamma_3z_3)}, 
\eeqn 
where $x_1$ is age, $x_2$ is weight before pregnancy,
$z_1$ is an indicator for being a smoker, whereas 
$z_2$ and $z_3$ are indicators for belonging to certain ethnic 
groups. The dataset in question comprises 189 mothers
and babies, with these five covariates having been recorded 
(along with yet others; see \citet[Ch.~2]{ClaeskensHjort08} 
for further discussion). The eight models correspond
to pushing the `open' covariates $z_1,z_2,z_3$ in and out 
of the logistic regression structure, while $x_1,x_2$ 
are `protected' covariates. The plot shows the point estimates
$\hatt p$ for the 8 different submodels on the vertical axis 
and {\it root-FIC scores} on the horizontal axis. 
These are estimated risks, i.e.~estimates of 
root-mean-squared-errors. Crucially, 
the FIC scores do not merely assess the standard deviation 
of estimators, but also take the potential biases on board, 
from using smaller models. 

Using the FIC ranking, as summarised both in the {\it FIC table} 
given in Table \ref{table:table1} and the FIC plot, 
therefore, we learn that submodels 000 and 010 are the best
(where e.g.~`010' indicates the model with $z_2$ on board
but without $z_1$ and $z_3$, etc.), 
associated with point estimates 0.282 and 0.259, 
whereas submodels 100 and 011 are the ostensibly worst,
with rather less precise point estimates 0.368 and 0.226.
Again, `best' and `worst' means as gauged by precision
of these 8 estimates of the same quantity. 
Importantly, the FIC machinery, as briefly explained 
here, with more details in later sections, 
can be used for each new woman, with different
`best models' for different strata of women, and
it may be used for handling different and even 
quite complicated focus parameters. 
In particular, if Mrs.~Jones had not been a smoker,
so that her $z_1=1$ would rather have been a $z_1=0$, 
we run our programmes to produce a FIC table and a FIC plot
for her, and learn that the submodel ranking is very different.
Then 111 and 101 are the best and 001 and 000 the worst; 
also, the $\hatt p$ estimates of her having a baby
with small birthweight are significantly smaller. 

\begin{table}[h]
\small
\begin{center}
% \begin{tabular}{ccc|cc|cc|cc|cc|cc} 
\begin{tabular}{c | c | c |  c | c | c | c}
% \toprule
   &in-or-out & $\hatt p$ & stdev & bias & root-FIC & rank \\ 
\midrule 
 1 & 0 0 0    &  0.282 & 0.039 & 0.000 & 0.039 & 1 \\
 2 & 1 0 0    &  0.368 & 0.055 & 0.061 & 0.082 & 7 \\
 3 & 0 1 0    &  0.259 & 0.042 & 0.000 & 0.042 & 2 \\
 4 & 0 0 1    &  0.267 & 0.048 & 0.000 & 0.048 & 3 \\
 5 & 1 1 0    &  0.342 & 0.057 & 0.037 & 0.068 & 5 \\
 6 & 1 0 1    &  0.351 & 0.056 & 0.045 & 0.072 & 6 \\
 7 & 0 1 1    &  0.226 & 0.054 & 0.063 & 0.083 & 8 \\
 8 & 1 1 1    &  0.303 & 0.060 & 0.000 & 0.060 & 4 \\
% \bottomrule
\end{tabular}
\end{center}
\caption{FIC table for Mrs.~Jones: 
there are $2^3=8$ submodels, with absence-presence of 
$z_1,z_2,z_3$ indicated with 0 and 1 in column 2, followed by 
estimates $\hatt p$, estimated standard deviation, 
% $\hatt\tau_S/\rootn$, 
estimated absolute bias,  
% $(\hatt\bsq_S)^{1/2}/\rootn$, 
the root-FIC score, which is also the Pythagorean
combination of the stdev and the bias, and the model rank. 
The numbers are computed with formulae of Section \ref{section:basics}.}
\label{table:table1}
\end{table}

The FIC apparatus, initiated and developed in 
\citet{ClaeskensHjort03}, \citet{HjortClaeskens03a}, 
\citet{ClaeskensHjort08}, has led to quite a rich
literature; see comments at the end of this section. 
FIC analyses have different forms of output, 
qua FIC tables (listing the best candidate models, 
along with estimates and root-FIC scores,
perhaps supplemented with more information) 
and FIC plots. The general setup involves a selected quantity
of particular interest, say $\mu$, called the focus 
parameter, and various candidate models, say $S$,
leading to a collection of estimators $\hatt\mu_S$. 
These carry root-mean-squared-errors $\rmse_S$,
and the root-FIC scores are estimates of these root-risks.
The FIC plot displays 
\beq
\label{eq:ficplot}
(\fic_S^{1/2},\hatt\mu_S)=(\hatt\mse_S^{1/2},\hatt\mu_S)  
   \quad {\rm for\ all\ candidate\ models\ } S,
\eeq 
as with Figure \ref{figure:fig14}. 

The present paper concerns {\it going beyond} such FIC plots,
investigating the precision of each displayed point.
The point estimates $\hatt\mu_S$ carry uncertainty,
as do the FIC scores. A more elaborate version 
of the FIC plot can therefore display the uncertainty
involved, in both the vertical and horizontal directions. 
This aids the statistician in seeing whether 
good models are `clear winners' or not, and 
whether the ostensibly best estimates are 
genuinely more accurate than others. In various
concrete examples one also observes that a few 
candidate models appear to be better than the rest.
The methodology of our paper makes it possible
to assess to which extent the implied differences 
in FIC scores are significant. Such insights
lead also to model averaging strategies with 
weights given precisely to the best models for the
given estimation purpose. 

Our paper proceeds as follows. 
In Section \ref{section:basics} we give the 
required mathematical background, involving 
both the basic notation necessary and the key theorems
about joint convergence of classes of candidate 
model based estimators. These results also drive
the development of confidence distributions 
for FIC scores, in Section \ref{section:CDforfic}.
Flowing from these results again is also a new 
variant for the FIC, which we call the quantile-FIC,
where each root mean squared error quantity (rmse) 
is naturally estimated using an appropriate quantile 
in the associated confidence distribution.
A special case is the median-FIC; 
details are given in Section \ref{section:medianfic}. 
With such results on board, 
Section \ref{section:modelaveraging} then involves
constructions of median-FIC driven weights 
for model averaging operations, where we also 
give a precise large-sample description of 
the implied model averaging estimators. 
In Section \ref{section:comparison} we address 
performance and comparison issues, studying 
relevant aspects of how well different schemes
behave, from post-FIC to model averaging estimators. 
It is in particular seen that the post-median-FIC
estimators have certain advantages over post-AIC
schemes. 
To display how our new CD-FIC based methods work
in a setup with considerably more candidate models
at play than with the $2^3=8$ models used for
Mrs.~Jones above, a multi-regression Poisson setup
is worked through in Section \ref{section:applications},
involving abundance of bird species for 73 British 
and Irish islands. Then we sum up various salient points
in our discussion Section \ref{section:discussion},
and round off our paper with a list of concluding
remarks, some pointing to further research, 
in Section \ref{section:concluding}. 

We end our introduction section by commenting briefly
on other relevant work, first on the FIC front 
and then on model averaging. 
Setting up FIC schemes involves finding good approximations
to mse quantities, and then constructing estimators 
for these. This pans out differently in different 
classes of models, and sometimes requires lengthy 
separate efforts, depending also on the type of 
interest parameter in focus. \citet{ClaeskensHjort08} 
cover a broad range of general i.i.d.~and regression models, 
using local neighbourhoods methodology. 
Later extensions include 
\citet{Claeskensetal07} for time series models, 
\cite{GueuningClaeskens18} for highdimensional setups,
\citet{HjortClaeskens06} and \citet{Hjort08Vonta} 
for semiparametric and nonparametric survival regression models,
\citet{ZhangLiang11} for generalised additive models,
\citet{Zhangetal12} for tobit models,
\citet*{KoHjort19} for copulae with two-stage estimation methods.
Recent methodological extensions and advances 
also include setups centred on a fixed wide model, 
with large-sample approximations not depending on 
the local asymptotics methods; 
see \citet*{JullumHjort17, JullumHjort19, 
ClaeskensCunenHjort19}, along with \citet*{CunenWalloeHjort19}
for linear mixed models. There is a growing list of 
application domains where FIC is finding practical 
and context-relevant use, such as 
finance and economics \citep{Brownless08, BehlDette12}, 
peace research and political science \citep*{CunenHjortNygaard20}, 
sociology \citep{Zhangetal12},
marine science \citep*{HermansenHjortKjesbu16}, etc.
There is similarly a rapidly expanding literature 
on frequentist model averaging procedures, as partly
contrasted with Bayesian versions; perspectives
for the latter are summarised in \citet{Hoetingetal99}.   
A broad framework for frequentist averaging methods
is developed in \citet{HjortClaeskens03a, ClaeskensHjort08},
including precise large-sample descriptions for 
how such schemes actually perform. 
\citet{WangZhangZou09} give a broad review.  
In econometrics, \citet{Hansen07} studies model averaging 
for least squares procedures, and \citet{Magnusetal09}
compare frequentist and Bayesian averaging methods.
Optimal weights are studied in \citet{Liangetal11}. 
The book chapter \citet{Chanetal20} discusses 
optimal averaging schemes for forecasting,
where a certain phenomenon is that simpler 
weighting methods sometimes perform better than 
those involving extra layers of estimation to
get closer to envisaged optimal weights. 
 
\section{The basic setup and the FIC}
\label{section:basics} 

In this section we give the basic theoretical background 
and main results behind the FIC plots (\ref{eq:ficplot}). 
It is convenient to describe the i.i.d.~setup first, 
and to describe a canonical limit experiment with the 
required basic quantities following from the relevant
assumptions. We then briefly explain how the apparatus 
can be extended also to general regression models, 
where it also turns out that the limit experiment 
is of exactly the same type, only with somewhat more
complex mechanisms lying behind the key ingredients. 
The key results described in this section are behind 
the FIC plots and the FIC tables, such as 
Figure \ref{figure:fig14} and Table \ref{table:table1}, 
and will also be used in later sections to derive 
confidence distributions for risks. 

\subsection{The i.i.d.~setup.}
\label{subsection:iid}

Suppose we have independent and identically distributed observations,
say $y_1,\ldots,y_n$. Under study is a collection of candidate
models, ranging from a well-defined narrow model,
parametrised as $f_\narr(y,\theta)$ with 
$\theta=(\theta_1,\ldots,\theta_p)$ of dimension $p$,
to a wide model, parametrised as $f(y,\theta,\gamma)$,
with certain extra parameters 
$\gamma=(\gamma_1,\ldots,\gamma_q)$, signifying model
extensions in different directions. The narrow model
is assumed to be an inner point in the wider model,
in the sense of $f_\narr(y,\theta)$ being equal to 
$f(y,\theta,\gamma_0)$ for an inner parameter point $\gamma_0$.
There is consequently a total of $2^q$ candidate models,
corresponding to setting $\gamma_j$ parameters equal to 
or not equal to their null values $\gamma_{0,j}$,
for $j=1,\ldots,q$. In the regression framework 
studied below this would typically correspond to 
taking covariates in and out of the wide model. 

Assume now that a parameter $\mu$ is to be estimated, 
with a clear statistical interpretation across candidate
models. It may in particular be expressed as $\mu=\mu(\theta,\gamma)$
in the wide model. We may then consider $2^q$ different candidate
estimators, say $\hatt\mu_S$ based on the submodel $S$,
with $S$ a subset of $\{1,\ldots,q\}$, corresponding 
to the model having $\gamma_j$ as a parameter in the model
when $j\in S$ but with $\gamma_j$ set to their null 
values $\gamma_{0,j}$ for $j\notin S$. Carrying out 
maximum likelihood (ML) estimation in model $S$ means maximising
the log-likelihood function 
$\ell_{n,S}(\theta,\gamma_S)=\sumin\log f(y_i,\theta,\gamma_S,\gamma_{0,S^c})$, 
with $\gamma_S$ notation for the collection of $\gamma_j$
with $j\in S$, and similarly for $\gamma_{0,S^c}$ 
with the complement set. With $(\hatt\theta_S,\hatt\gamma_S)$
the ML estimators for submodel $S$, this leads to a collection 
of candidate estimators 
\beqn
\hatt\mu_S=\mu(\hatt\theta_S,\hatt\gamma_S,\gamma_{0,S^c}) 
   \quad {\rm for\ }S\in\{1,\ldots,q\}. 
\eeqn 
In particular we have $\hatt\mu_\narr=\mu(\hatt\theta_\narr,\gamma_0)$
and $\hatt\mu_\wide=\mu(\hatt\theta_\wide,\hatt\gamma_\wide)$,
with ML estimation carried out in respectively the narrow 
$p$-dimensional and the wide $(p+q)$-dimensional models. 

To understand the behaviour of all these candidate estimators,
and to develop theory and methods for sorting them through,
aiming at finding the best, we now present a `master theorem',
from \citet{HjortClaeskens03a}, \citet[Chs.~5, 6]{ClaeskensHjort08}.  
We work inside a system of local neighbourhoods, where 
the real data-generating mechanism underlying our observations is 
\beq
\label{eq:ftrue}
f_\true(y)=f(y,\theta_0,\gamma_0+\delta/\rootn),  
\eeq
with some unknown $\delta=\rootn(\gamma-\gamma_0)$, 
seen as a local model extension parameter;
in particular, the true focus parameter becomes 
$\mu_\true=\mu(\theta_0,\gamma_0+\delta/\rootn)$. 
A few key quantities now need proper definition. We start with 
the Fisher information matrix with inverse, of the wide model,
but computed at the null model: 
\beq
\label{eq:JandJinverse}
J=\begin{pmatrix} J_{00} &J_{01} \\ J_{10} & J_{11} \end{pmatrix}
   \quadandquad
  J^{-1}=\begin{pmatrix} J^{00} &J^{01} \\ J^{10} & J^{11} \end{pmatrix}. 
\eeq
Here blocks $J_{00}$ and $J^{00}$ are of size $p\times p$, etc. 
The $q\times q$ matrix 
\beqn
Q=J^{11}=(J_{11}-J_{10}J_{00}^{-1}J_{01})^{-1}
\eeqn 
serves a vital role. So do also 
\beq
\label{eq:omegaandtau0}
\omega=J_{10}J_{00}^{-1}\dellone-\delltwo 
   \quadandquad
  \tau_0^2=(\dellone)^\tr J_{00}^{-1}\dellone, 
\eeq 
with partial derivatives evaluated at the null model, 
and with these quantities varying from focus parameter 
to focus parameter. Finally we need to introduce the 
$q\times q$ matrices 
\beqn
G_S=\pi_S^\tr Q_S\pi_S Q^{-1}, \quad {\rm with} \quad 
   Q_S=(\pi_S Q^{-1}\pi_S^\tr)^{-1}. 
\eeqn 
We have $G_\narr=0$ and $G_\wide=I$, 
the $q\times q$ identity matrix, 
and note that $\Tr(G_S)=|S|$, the number of elements in $S$. 

The master theorems driving much of the FIC and 
related theory are now as follows. First, 
\beq
\label{eq:Dn}
D_n=\rootn(\hatt\gamma_\wide-\gamma_0)\arr_d D\sim\N_q(\delta,Q), 
\eeq  
and, secondly, 
\beq
\label{eq:master1}
\rootn(\hatt\mu_S-\mu_\true)\arr_d
   \Lambda_S=\Lambda_0+\omega^\tr(\delta-G_SD) 
   \quad {\rm for\ each\ }S\in\{1,\ldots,q\}. 
\eeq 
Here $\Lambda_0\sim\N(0,\tau_0^2)$, for the $\tau_0$
given above, and $\Lambda_0$ and $D$ are 
independent. This implies that the limit in (\ref{eq:master1})
is normal, and we can read off its bias $\omega^\tr(I-G_S)\delta$ 
and variance $\tau_0^2+\omega^\tr G_SQG_S^\tr\omega$. 
The risk or mean squared error for this limit distribution is hence 
\beq
\label{eq:mseS}
\mse_S=\E\,\Lambda_S^2
      =\tau_0^2+\omega^\tr G_SQG_S^\tr\omega+\{\omega^\tr(I-G_S)\delta\}^2
   =\var_S+\bsq_S, 
\eeq 
say, in the usual fashion a sum of a variance part $\var_S$
and a squared bias part $\bsq_S$. With a slim $S$, there
are many zeros in $G_S$, leading to small variance 
but potentially a bigger bias; with a fatter $S$, $G_S$ 
becomes closer to the identity matrix $I$, 
yielding bigger variance but a smaller bias. 

The essence of the Focused Information Criterion (FIC), 
developed in \citet{ClaeskensHjort03, ClaeskensHjort08} 
and later extended in various directions and to more general 
contexts and model classes, is to estimate each $\mse_S$
from the data. This leads to a full ranking of all candidate
models, from the best, meaning smallest estimate of risk,
to the worst, meaning largest estimates of risk. 
Briefly, we start by putting up FIC formulae 
for the limit experiment, where all quantities
$\tau_0,Q,G_S,\omega$ are known (thanks to consistent
estimators for these, see below), but where $\delta$ is not, 
as we can only rely on the information $D\sim\N_q(\delta,Q)$ 
from (\ref{eq:Dn}). Noting that $\E\,DD^\tr=\delta\delta^\tr+Q$, 
which also means that using $(c^\tr D)^2$ to estimate 
a squared linear combination parameter $(c^\tr\delta)^2$
means overshooting with expected amount $c^\tr Qc$,  
there are actually two natural versions here, namely 
\beq
\label{eq:ficficlimit}
\begin{array}{rcl}
\!\!\!\!\!\!\fic^u&=&\displaystyle
\var_S+\hatt\bsq_S
   =\tau_0^2+\omega^\tr G_SQG_S^\tr\omega+
   \omega^\tr(I-G_S)(DD^\tr-Q)(I-G_S)^\tr\omega, \\
\!\!\!\!\!\!\fic^t&=&\displaystyle
\var_S+\tilda\bsq_S
   =\tau_0^2+\omega^\tr G_SQG_S^\tr\omega+
   \max\{\omega^\tr(I-G_S)(DD^\tr-Q)(I-G_S)^\tr\omega,0\}, 
\end{array}
\eeq  
corresponding to a directly unbiased estimator and 
its truncated-to-zero version for the squared bias.
That the first estimator for squared bias is negative
means that the event 
\beqn
%% \label{eq:truncationdecider}
\{\omega^\tr(I-G_S)D\}^2<\omega^\tr(I-G_S)Q(I-G_S)^\tr\omega, 
\eeqn    
is taking place, which happens quite frequently if 
$\delta$ is close to zero, in fact with probability up to 
$\Pr\{\chi^2_1\le1\}=0.683$, if $\delta=0$, but is growing
less likely when $\delta$ is moving away from zero. 

For actual data one plugs in consistent estimators 
$\hatt\tau_0,\hatt Q,\hatt G_S,\hatt\omega$ for the
relevant quantities, to be given below, 
%% see again Section \ref{section:basicfic}, 
and $D_n$ of (\ref{eq:Dn}) for $\delta$.
This leads to FIC scores 
\beq
\label{eq:ficficdata}
\begin{array}{rcl}
\fic^u&=&\displaystyle
   \hatt\tau_0^2+\hatt\omega^\tr \hatt G_S\hatt Q\hatt G_S^\tr\hatt\omega
   +\{\hatt\omega^\tr(I-\hatt G_S)D_n\}^2
   -\hatt\omega^\tr(I-\hatt G_S)\hatt Q(I-\hatt G_S)^\tr\hatt\omega, \\  
\fic^t&=&\displaystyle
 \hatt\tau_0^2+\hatt\omega^\tr \hatt G_S\hatt Q\hatt G_S^\tr\hatt \omega
   +\max\bigl[\{\hatt\omega^\tr(I-\hatt G_S)D_n\}^2
   -\hatt\omega^\tr(I-\hatt G_S)\hatt Q(I-\hatt G_S)^\tr\hatt\omega,0\bigr]. \\  
\end{array}
\eeq  
Note from (\ref{eq:master1}) that these are estimators 
of the limiting risk, where $\hatt\mu_S-\mu_\true$ has 
been multiplied with $\rootn$. Most often it is therefore
better, regarding reading of tables and interpretation of 
FIC plots, to transform the above scores to say 
\beq
\label{eq:rootfic}
\rootfic^u=(\fic^u)^{1/2}/\rootn 
\quadandquad
\rootfic^t=(\fic^t)^{1/2}/\rootn.  
\eeq 
We consider the truncated version a good default 
choice, since it avoids having negative estimates
of squared biases, and this choice has indeed
been used for Mrs.~Jones and her FIC plot in Figure \ref{figure:fig14}
and FIC table in Table \ref{table:table1}. 
The consistent estimators in question are computed 
as follows. From ML analysis in the wide model, maximising 
$\ell_{n,\wide}(\theta,\gamma)$, we compute the normalised 
Hessian matrix at this ML position, say 
$\hatt\eta_\wide=(\hatt\theta_\wide,\hatt\gamma_\wide)$, 
\beqn
\hatt J_\wide=-n^{-1}{\dell^2\ell_{n,\wide}(\hatt\eta_\wide)
   \over \dell\eta\dell\eta^\tr}, 
\eeqn 
of size $(p+q)\times(p+q)$. This is a consistent estimator
for $J$ of (\ref{eq:JandJinverse}) under the 
assumed sequence of data-generating mechanisms 
(\ref{eq:ftrue}), under mild conditions; see 
\citet[Ch.~6]{ClaeskensHjort08}. Inverting this matrix 
and reading off its lower right block leads to 
$\hatt Q=\hatt J^{11}$, consistent for $Q$. 
Finally $\hatt\omega$ and $\hatt\tau_0$ are defined
by plugging in relevant blocks of $\hatt J_\wide$
in (\ref{eq:omegaandtau0}), along with partial derivatives
of $\mu(\theta,\gamma)$, computed at the ML position
$(\hatt\theta_\wide,\hatt\gamma_\wide)$. 
There are in fact a few alternatives here, regarding
estimation of $J$ and $\omega$, but these do not 
affect the basic asymptotics; see 
\citet[Ch.~6, 7]{ClaeskensHjort08} for further discussion. 

For simplicity we have chosen not to overburden the notation
here, with one name for FIC in the limit experiment, 
as in (\ref{eq:ficficlimit}), and another for FIC 
with real data, as in (\ref{eq:ficficdata}); it is in
each case clear from the context what is what. 

\subsection{Extension to regression models.}
\label{subsection:regressions}

As demonstrated in \citet{ClaeskensHjort03, ClaeskensHjort08},
the theory briefly reviewed above for the i.i.d.~setup
can with the required extra efforts be lifted to the 
framework of regression models. Data are then of the form 
$(x_i,y_i)$, with $x_i$ a covariate vector and $y_i$ the response.
The natural setup becomes that of a wide regression model
with densities $f(y_i\midd x_i,\theta,\gamma)$, 
featuring a narrow model parameter $\theta$ of size $p$
and an extra $\gamma$ parameter of size $q$,
and where a null value $\gamma=\gamma_0$ yields the narrow model. 
Again using $\gamma=\gamma_0+\delta/\rootn$ as the 
natural framework of local asymptotics, there are under 
mild Lindeberg conditions clear limiting normality results 
for all submodel based estimators, etc., though involving 
somewhat more complex notation than for the i.i.d.~case 
when it comes to key quantities $Q,\omega,G_S$. 

It is however simplest to develop our extended 
CD-FIC theory for the i.i.d.~case, which we make our task below.
For each method and result reached below there is a 
natural extension to the case of regression models. 
This is illustrated in Section \ref{section:applications}
for a class of Poisson regression models applied 
to a study of bird species abundance.  

\section{Confidence distributions for FIC scores} 
\label{section:CDforfic}

The FIC scores of (\ref{eq:ficficlimit}) are estimators 
of the $\mse_S$ quantities (\ref{eq:mseS}), defined
in the limit experiment where $D\sim\N_q(\delta,Q)$
and the other key quantities are known. Similarly,
the $\rootfic$ scores of (\ref{eq:rootfic}) 
are estimating the genuine $\rmse_S$, the root-mse
for the real estimators $\hatt\mu_S$.    
But the FIC scores carry their own uncertainty, 
which we address in this section through constructing 
confidence distributions for the estimated quantities. 

As in Section \ref{section:basics} we start working
out matters in the clear limit experiment, and 
then insert consistent estimators when engaged 
with real data. A brief prelude to explain what 
will take place is as follows:
Suppose a single $X$ is observed from a $\N(\eta,1)$,
and that inference is needed for the parameter $\phi=\eta^2$.
Since $X^2$ is a noncentral chi-squared, with 1 degree of 
freedom and noncentrality parameter $\eta^2$, 
which we write as $X^2\sim\chi^2_1(\eta^2)$, 
we can build the function 
\beqn
C(\phi)=C(\phi,x_\obs)=\Pr_\eta\{X^2\ge x_\obs^2\}
   =1-\Gamma_1(x_\obs^2,\phi), 
\eeqn 
with $\Gamma_1(\cdot,\phi)$ the cumulative distribution
function for the $\chi^2_1(\phi)$. Here $x_\obs$ is the
observed value of the random $X$. The $C(\phi,x_\obs)$ is 
a cumulative distribution function in $\phi$, for 
the observed $x_\obs$, with the property that 
for each $\eta$, when $X$ comes from the data model $\N(\eta,1)$,
then $C(\phi,X)$ has the uniform distribution: 
\beqn
\Pr_\eta\{C(\phi,X)\le\alpha\}=\alpha
   \quad {\rm for\ each\ }\alpha. 
\eeqn  
In other words, $C(\phi,x)$ defines a full 
and exact confidence distribution (CD), see 
\citet{SchwederHjort16,HjortSchweder18}, 
and confidence intervals can be read off from 
$\{\phi\colon C(\phi,x_\obs)\le\alpha\}$. 
Note that this CD has a pointmass at zero, 
$C(0,x_\obs)=1-\Gamma_1(x_\obs^2)$, involving the standard 
chi-squared cumulative $\Gamma_1(\cdot)=\Gamma_1(\cdot,0)$. 
Thus confidence intervals for $\phi=\eta^2$ could very
well start at zero. This CD is the optimal one, 
in this situation, cf.~\citet[Ch.~6]{SchwederHjort16}. 

Going back to the $\mse_S$ of (\ref{eq:mseS}), write 
\beqn
\mse_S 
=\tau_0^2+\omega^\tr G_SQG_S^\tr\omega+\{\omega^\tr (I-G_S)\delta\}^2 
=\tau_S^2+\sigma_S^2\Bigl\{{\omega^\tr(I-G_S)\delta\over \sigma_S}\Bigr\}^2, 
\eeqn  
with 
\beqn
\tau_S^2=\tau_0^2+\omega^\tr G_SQG_S^\tr\omega
   \quadandquad
\sigma_S^2=\omega^\tr(I-G_S)Q(I-G_S)^\tr. 
\eeqn
Here $\tau_S^2$ is the limiting variance of $\rootn\hatt\mu_S$, 
which is smaller with fewer elements in $S$ 
and correspondingly larger with more elements in $S$. 
Also, $\sigma_S^2$ is the variance of $\omega^\tr(I-G_S)D$, 
i.e.~of the estimate of the bias $\omega^\tr(I-G_S)\delta$. 
Write for clarity 
$X_S=\omega^\tr(I-G_S)D/\sigma_S$, which has a 
$\N(\eta_S,1)$ distribution, with $\eta_S=\omega^\tr(I-G_S)\delta/\sigma_S$. 
Since quantities $\tau_0,\omega,Q,G_S$ are known, 
in the limit experiment, the arguments above lead to the CD 
\beq
\label{eq:CSmseS}
\begin{array}{rcl}
C_S(\mse_S)
&=&\displaystyle
\Pr_\delta\{\tau_S^2+\sigma_S^2X_S^2
   \ge\tau_S^2+\sigma_S^2X_{S,\obs}^2\} \\
&=&\displaystyle
1-\Gamma_1\Bigl({\{\omega^\tr(I-G_S)D_\obs\}^2\over \sigma_S^2},
   {\mse_S-\tau_S^2\over \sigma_S^2}\Bigr)
   \quad {\rm for\ }\mse_S\ge\tau_S^2. 
\end{array}
\eeq
It starts at position $\tau_S^2$, the minimal possible
value for $\mse_S$, with pointmass there of size 
$C_S(\tau_S^2)=1-\Gamma_1(\{\omega^\tr(I-G_S)D_\obs/\sigma_S\}^2)$. 

The narrow model, with $S=\emptyset$ and $G_\narr=0$, 
has the smallest $\tau_S$, namely $\tau_0$, but also 
the largest $\sigma_S$, with  
%% and {\color{red} the most uncertain CD 
%% [xx not necessarily, because it can start at a high point xx]}, with  
\beqn
C_\narr(\mse_\narr)
=1-\Gamma_1\Bigl( {(\omega^\tr D_\obs)^2\over \omega^\tr Q\omega},
   {\mse_\narr-\tau_0^2\over \omega^\tr Q\omega}\Bigr)
   \quad {\rm for\ }\mse_\narr\ge\tau_0^2. 
\eeqn 
On the other side of the spectrum of candidate models, 
the widest model has $G_\wide=I$, the $\mse_\wide$
is the constant $\tau_0^2+\omega^\tr Q\omega$ with no
additional uncertainty, in this framework of the limit
experiment, and the $C_\wide(\mse_\wide)$ is simply 
a full pointmass 1 at that position. 

For a real dataset, we estimate the required quantities 
consistently, as per Section \ref{section:basics}, 
and with $D_n=\rootn(\hatt\gamma_\wide-\gamma_0)$
of (\ref{eq:Dn}) for $D$. Translating and transforming
also to the real root-mse scale of 
\beqn
\rho_S=\rmse_S/\rootn, \quad {\rm for\ } \hatt\mu_S-\mu_\true, 
\eeqn 
we reach the real-data based CD 
\beq
\label{eq:CDforrho}
C_S^*(\rho_S)
=1-\Gamma_1\Bigl({\{\hatt\omega^\tr(I-\hatt G_S)D_n\}^2\over \hatt\sigma_S^2},
   {n\rho_S^2-\hatt\tau_S^2\over \hatt\sigma_S^2}\Bigr)
   \quad {\rm for\ }\rho_S\ge\hatt\tau_S/\rootn.  
\eeq  
Here 
$\hatt\tau_S^2=\hatt\tau_0^2+\hatt\omega^\tr 
   \hatt G_S\hatt Q\hatt G_S^\tr\hatt\omega$ 
and 
$\hatt\sigma_S^2=\hatt\omega^\tr(I-\hatt G_S)\hatt Q(I-\hatt G_S)^\tr\hatt\omega$,
and the CD starts with the pointmass 
$C_S^*(\hatt\tau_S/\rootn)
=1-\Gamma_1(\{\hatt\omega^\tr(I-\hatt G_S)D_n/\hatt\sigma_S\}^2)$
at its minimal position $\hatt\tau_S/\rootn$.  
The CD $C_S^*(\rho_S)$ is large-sample correct, in the 
sense that for any given position in the parameter space,
its distribution converges to that of the uniform
as sample size increases. Thus 
$\{\rho_S\colon C_S^*(\rho_S)\le\alpha\}$ defines 
a confidence interval for $\rho_S$, with coverage 
converging to $\alpha$. 

\begin{figure}[h]
\centering
\includegraphics[scale=0.5]{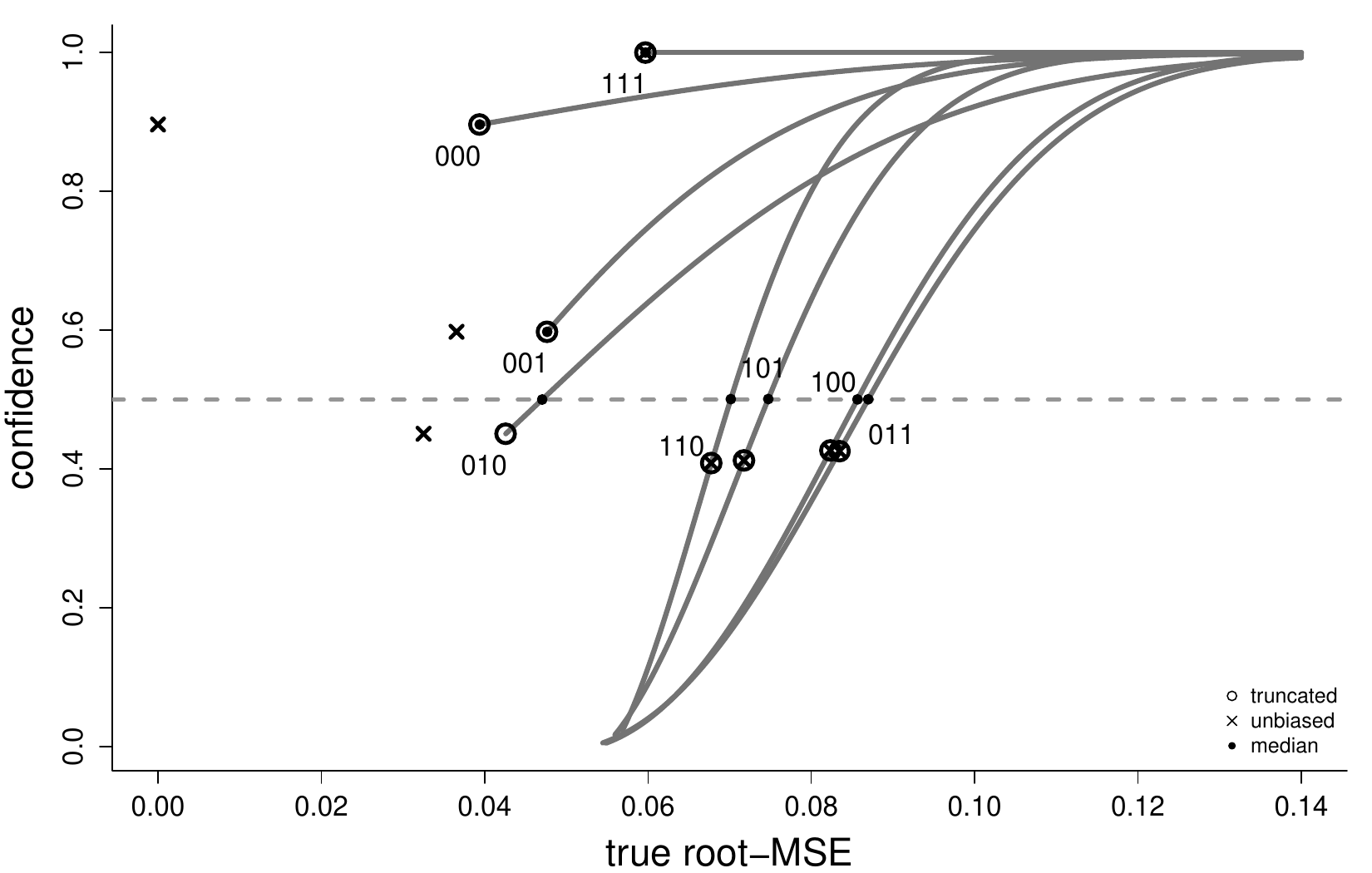}
\caption{Confidence distributions for the true root-mse 
values of the eight submodels in the Mrs.~Jones example.}
\label{figure:fig21}
\end{figure}

In Figure \ref{figure:fig21} confidence distributions 
are displayed for the eight true root-mse values pertaining 
to the eight submodels in the  Mrs.~Jones example of our introduction. 
Clearly, several of the CDs have pointmasses well above zero. 
% The wide model simply has a pointmass in 1, as will always be 
% the case in this framework. 
Also displayed in the figure are 
three root-FIC scores of different type: the already mentioned 
$\fic^u$ and $\fic^t$, along with the median-FIC 
which we come to in the next section. The unbiased estimator 
$\fic^u$ can for some models be considerably smaller 
than $\fic^t$; indeed it has the value zero for the narrow model 000. 
% (note that there is a truncation involved here too, 
% to avoid errors when trying to take the square-root of a negative number). 
The models with smaller $\fic^u$ than $\fic^t$ have negative 
squared bias estimates, 
i.e.~$\{\hatt\omega^\tr(I-\hatt G_S)D_n\}^2 < \hatt\sigma_S^2$, 
then the ratio $\{\hatt\omega^\tr(I-\hatt G_S)D_n\}^2/\hatt\sigma_S^2$ 
inside $\Gamma_1(\cdot)$ will be smaller than 1, 
which leads to the corresponding CDs starting 
with a pointmass higher than $0.3173=1-\Gamma_1(1)$.

\begin{figure}[h]
\centering
\includegraphics[scale=0.5]{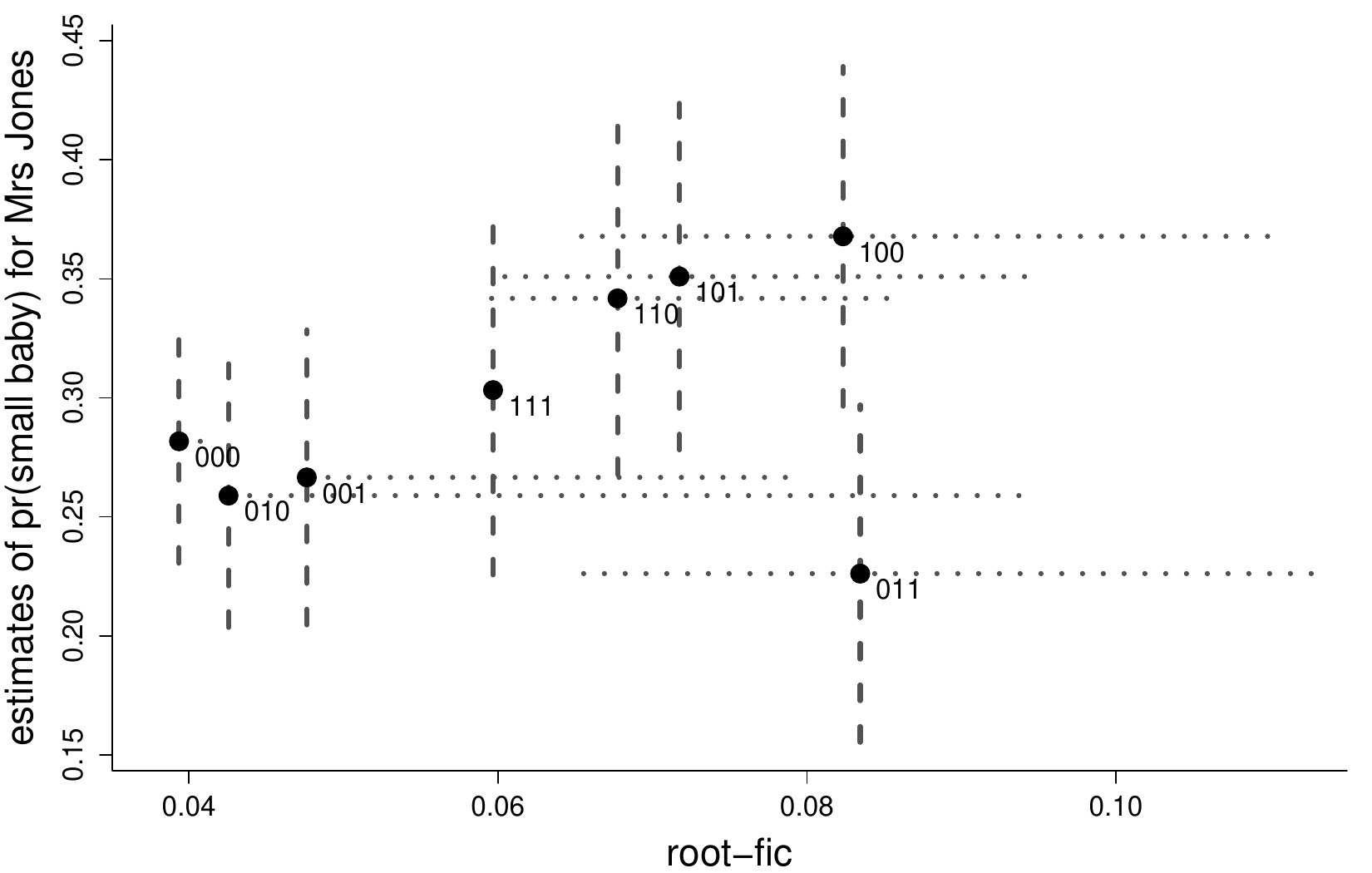}
\caption{FIC plot with associated uncertainty for the $2^3=8$ 
models for estimating the probability of having a small child, 
for Mrs.~Jones (white, age 25, 60 kg, smoker). 
The uncertainty is represented by 80\% confidence intervals. 
The intervals for the root-FIC score are read off from the 
confidence distributions in Figure \ref{figure:fig21}. 
The intervals for the focus parameter are based on 
the ordinary normal approximation with estimated 
variances taken from the variance part of the FIC calculations 
(see e.g.~Table \ref{table:table1}). Note that the points
here are the `ordinary' truncated FIC scores.}
\label{figure:fig22}
\end{figure}

In our first exposition of the case of Mrs.~Jones, 
Figure \ref{figure:fig14} gave eight point estimates
for the probability of her child-to-come having 
small birthweight, along with root-FIC scores. From the CDs in 
Figure \ref{figure:fig21} we can construct an updated
and statistically more informative FIC plot, 
namely Figure \ref{figure:fig22}, which provides 
accurate supplementary information regarding 
how precise these root-FIC scores are. 
The figure provides confidence intervals for both the 
root-FIC scores and the focus estimates. In particular, 
we see that the FIC score for the winning model 000 
appears to be very precise, and we may then select this model 
without many misgivings. The scores of the next best models 
010 and 001 appear to be rather more uncertain, 
and their intervals indicate that their underlying true rmse 
values are potentially much larger than what their 
root-FIC scores indicate.

\section{The median-FIC and quantile-FIC}
\label{section:medianfic}

As briefly pointed to in Section \ref{section:basics}, there are 
often two valid variations on the basic FIC,
when it comes to estimating the precise $\rmse_S$ quantities,
as in (\ref{eq:ficficlimit}) and (\ref{eq:ficficdata}).
The first utilises the unbiased risk estimator, involving
the possibility of having negative estimates for 
squared biases, whereas these are truncated up to zero
for the second version. 

Since the most natural way of assessing uncertainty
of these risk estimators is via CDs, as in 
Section \ref{section:CDforfic}, with confidence 
pointmasses at the smallest values, etc., a third 
version suggests itself, namely the median confidence
estimators. Generally, these have unbiasedness 
properties on the median scale, as opposed to on 
the expectation scale, and are discussed 
in \citet[Chs.~3, 4]{SchwederHjort16}. 
Thus consider the median-FIC, 
\beq
\label{eq:medfic}
\medfic_S=C_S^{-1}(\half)
   =\min\{\mse_S\colon C_S(\mse_S)\ge\half\}, 
\eeq 
defined for the limit experiment, via (\ref{eq:CSmseS}),
to be viewed as an alternative to $\fic_S^u$ and $\fic_S^t$
of (\ref{eq:ficficlimit}).  
For actual data, having estimated the required background 
quantities and also transformed to the scale of $\rho_S=\rmse_S/\rootn$, 
we use the CD $C_S^*(\rho_S)$ of (\ref{eq:CDforrho}), 
and infer the median-FIC score 
\beq
\label{eq:medficstar}
{\medfic}_S^*=(C_S^*)^{-1}(\half)
   =\min\{\rho_S\colon C_S^*(\rho_S)\ge\half\}. 
\eeq 
See Figure \ref{figure:fig21} where we display the 0.50 
confidence line and read off the corresponding medians. 

Considering the limit experiment case (\ref{eq:medfic}) first, 
we know that the CD $C_S(\mse_S)$ starts out at the minimal point $\tau_S^2$
with the pointmass $1-\Gamma_1(\{\omega^\tr(I-G_S)D_\obs/\sigma_S\}^2)$.
If this is already at least $\half$, which inspection shows 
is equivalent to $r_S=|\omega^\tr(I-G_S)D/\sigma_S|\le 0.6745$, 
then the median-FIC is equal to $\tau_S^2$. 
If that ratio is above 0.6745, however, then the median-FIC
is the numerical solution to 
\beqn
1-\Gamma_1\Bigl({\{\omega^\tr(I-G_S)D_\obs\}^2\over \sigma_S^2},
   {\mse_S-\tau_S^2\over \sigma_S^2}\Bigr)=\half, 
\eeqn 
viewed as an equation in $\mse_S>\tau_S^2$. 
Similarly, when computing the median-FIC for a given dataset, 
we see that if 
\beqn
\rootn|\hatt\omega^\tr(I-\hatt G_S)(\hatt\gamma_\wide-\gamma_0)/\hatt\sigma_S|
=\rootn\Big|{\hatt\omega^\tr(I-\hatt G_S)(\hatt\gamma_\wide-\gamma_0)
   \over \{\hatt\omega(I-\hatt G_S)\hatt Q(I-\hatt G_S)^\tr\hatt\omega\}^{1/2}}\Big|
\le 0.6745,
\eeqn 
then the median-FIC for $C_S^*(\rho_S)$ is 
equal to the minimum value $\hatt\tau_S/\rootn$,
and otherwise one solves $C_S^*(\rho_S)=\half$
numerically with a solution to the right of $\hatt\tau_S/\rootn$. 

Going back to the limit experiment framework again,
with $r_S=|\omega^\tr(I-G_S)D/\sigma_S|$
the relative size of the estimated bias versus its uncertainty, 
we have the following relations between the 
three different FIC scores. 
(i) If $r_S\le 0.675$, 
   then $\fic^m_S = \fic^t_S = \tau_S^2 > \fic^u_S$;
(ii) if $0.675 < r_S < 1$, 
   then $\fic^m_S \ge \fic^t_S = \tau_S^2 > \fic^u_S$;
(iii) if $r_S \ge 1$, 
   then $\fic^m_S \ge \fic^t_S = \fic^u_S  \ge  \tau_S^2$.
In particular, it is always the case that 
$\fic^m_S \geq \fic^t_S  \geq \fic^u_S$.
Since the three types of FIC scores are identical 
for the wide model, the three strategies can be understood 
as having increasing preference for selecting the wide model.
The unbiased-FIC generally gives smaller FIC scores 
to all models except the wide model, so it will therefore
have a smaller probability of selecting the wide. 
The median-FIC, on the other hand, 
typically gives larger FIC scores to the competing models, 
and is then more likely to select the wide model. 
The truncated-FIC lies somewhere between these two approaches. 
We will compare the three strategies in more detail 
in Section \ref{section:comparison}, where each strategy
is studied also in terms of the risk of the estimator 
which the FIC score selects.

In addition to the median confidence estimator 
associated with the CDs it is also fruitful 
to consider the slightly more general quantile-FIC, which is 
\beq
\label{eq:quantilefic}
\fic^q_S=C_S^{-1}(q)
   =\min\{\mse_S\colon C_S(\mse_S)\ge q\}, 
\eeq  
for any given $q\in(0,1)$. We learn in 
Section \ref{section:comparison} that quantile 
values smaller than 0.50 may be beneficial for 
estimating the squared bias parts when these 
are small to moderate. 

Similarly to our brief comments about the median-FIC score above, 
we may work out some of the relations between the previously 
existing FIC scores and the quantile-FIC score. We may 
for example study the specific choice of $q=0.25$. 
This score, denoted by $\fic^{0.25}$, will be equal to 
$\tau_S^2$ when $r_S\le 1.1503$. For larger $r_S$ values 
one needs to find the numerical solution of $C_S(\mse_S) = 0.25$. 
Naturally, $\fic^m_S \ge \fic^{0.25}_S$. 
Further, if $r_S\le 1$, then $\fic^{0.25}_S = \fic^t_S > \fic^u_S$, 
but if $r_S > 1$, then $\fic^t_S = \fic^u_S \ge \fic^{0.25}_S$. 
The lower-quartile-FIC will thus often be smaller than 
the previously existing FIC scores, as opposed to the 
median-FIC which will always be larger or equal, 
as we saw above. Since all the FIC scores are identical
for the wide model, this entails that $\fic^{0.25}_S$ 
will exhibit a preference for selecting smaller models. 
We will come back to these insights in the discussion section. 

\section{Model averaging}  
\label{section:modelaveraging}

\begin{figure}[h]
\centering
\includegraphics[scale=0.5]{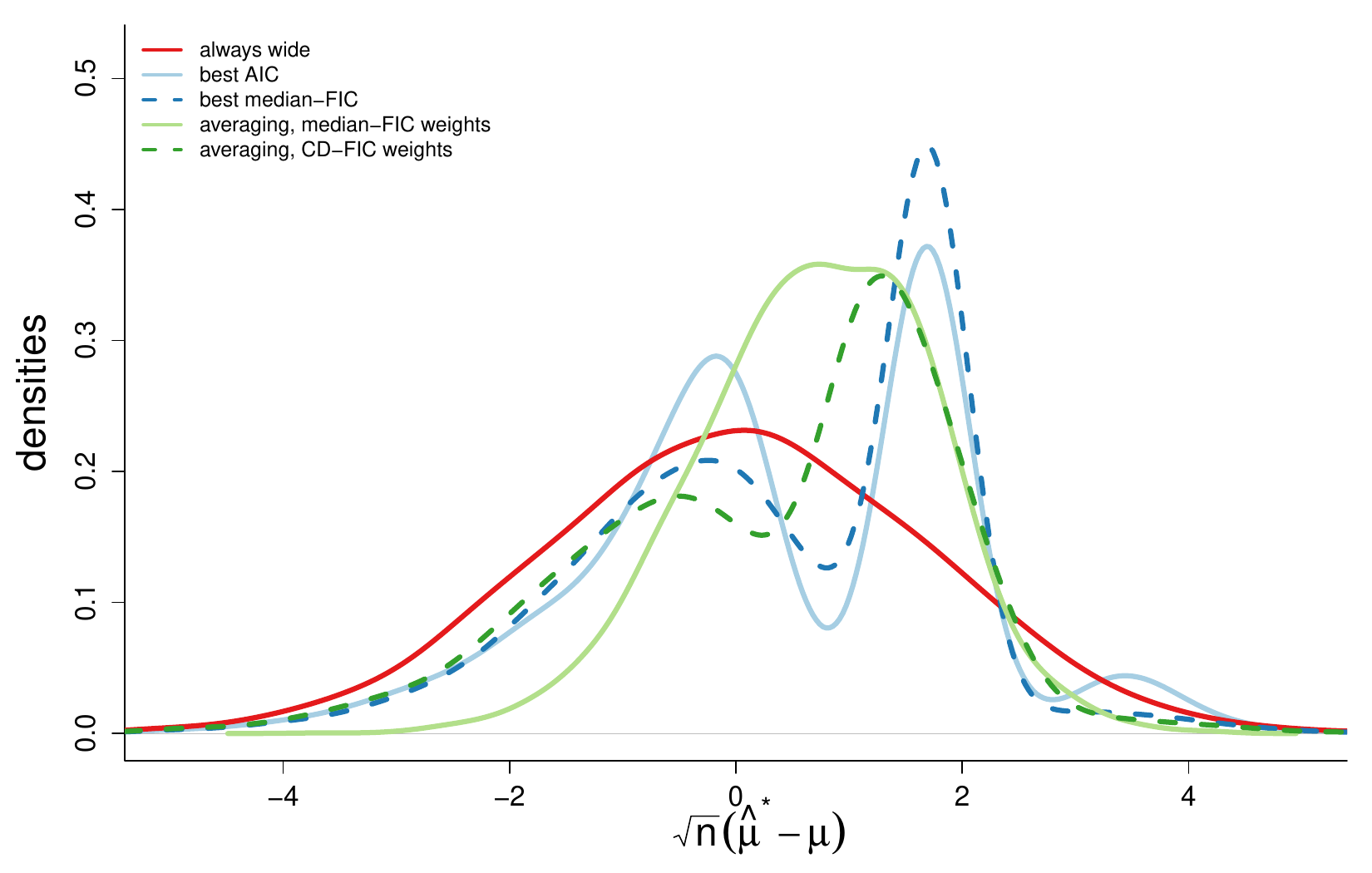}
\caption{Densities for limit distributions
of $\rootn(\hatt\mu^*-\mu_\true)$, for various 
choices of post-selection and model averaging $\hatt\mu^*$.}
\label{figure:fig16}
\end{figure}

Our FIC refinement investigations also invite new and 
focused model averaging schemes, where the weights
attached to the different candidate models are 
allowed to depend on the specific focus parameter 
under consideration. 
Consider model averaging estimators of the general form 
\beq
\label{eq:averaging}
\hatt\mu^*=\sum_S v_n(S\midd D_n)\hatt\mu_S, 
\eeq 
with weights depending on $D_n=\rootn(\hatt\gamma_\wide-\gamma_0)$
of (\ref{eq:Dn}), assumed to sum to 1, and 
with limits $v_n(S\midd D_n)\arr_d v(S\midd D)$. 
Coupling $D_n\arr_d D$ with 
\beqn
\rootn(\hatt\mu_S-\mu_\true)\arr_d
   \Lambda_S=\Lambda_0+\omega^\tr(\delta-G_SD) 
   \quad {\rm for\ each\ }S\in\{1,\ldots,q\} 
\eeqn 
of (\ref{eq:master1}), and utilising the joint 
limit distribution for the $2^q+1$ variables involved, 
a master theorem is reached in \citet[Ch.7]{ClaeskensHjort08}
of the form 
\beq
\label{eq:mastermaster}
\rootn(\hatt\mu^*-\mu_\true)
   \arr_d \Lambda_0+\omega^\tr\{\delta-\hatt\delta(D)\},
\quad {\rm where\ }
\hatt\delta(D)=\sum_S v(S\midd D)G_SD.  
\eeq 
This result is generalised to yet larger classes of 
model averaging strategies, including bagging
procedures, in \citet{Hjort14}, 

In the present context, a natural averaging estimator is 
as above, with weights of the form 
\beq
\label{eq:lambdaweights} 
v_n(S\midd D)=\exp(-\lambda\,\medfic_S)\Big/
   \sum_{S'}\exp(-\lambda\,\medfic_{S'}). 
\eeq 
The master theorem applies, which means we can read off
the accurate limit distribution for the median-FIC
based model averaging scheme in question.
We may also use different tuning parameters for different
models, i.e.~with weights proportional to 
$\exp(-\lambda_S\,\medfic_S)$, with appropriately 
selected $\lambda_S$. 
A general venue is to use the CDs for each model 
in order to set such model-specific $\lambda_S$ values.
One possibility is to evaluate all the CDs at the 
estimated rmse value of the widest model and then let 
\beq
\label{eq:coollambda}
\lambda_S = 1 / C^*_S(\fic_\wide^{1/2}/\rootn),
\eeq
see (\ref{eq:CDforrho}). For the wide model we have $\lambda_\wide=1$ 
but for the other models the $\lambda_S$ 
will have values above 1. The intuition is that 
dividing the FIC score with the confidence, 
evaluated at this specific point, will give higher weights 
to models where the FIC scores are more certain. This is the 
method we have employed for Figure \ref{figure:fig16}, 
for the model averaging scheme there denoted `CD-FIC weights'. 
%% with usual fic^t things, not fic^m? 

There are clearly several other model averaging schemes 
that may be considered based on the CDs for the FIC scores. 
One may e.g.~wish to use only models which have 
a high probability of having a rmse lower than 
a certain threshold, and then use a similar weighting scheme 
as above among the models with scores falling below 
this threshold. Again our master theorem (\ref{eq:mastermaster})
applies, with a precise description of the large-sample 
distributions of the ensuing model averaging estimators. 

In Figure \ref{figure:fig16} we present a brief illustration 
of different model selection and averaging schemes. The figure 
displays the limiting distribution densities 
of $\rootn(\hatt\mu^*-\mu_\true)$, 
from (\ref{eq:mastermaster}), for five different strategies. 
The densities are produced not by simulating from 
some given model with a high sample size, 
but from the exact limit distributions, 
by drawing from $\Lambda_0$ and $D$.
A sharper density around zero indicates that the strategy 
produces a more precise estimator than the others. The 
sharpness around zero may be assessed by computing the 
limiting mse of each $\rootn(\hatt\mu^* - \mu_\true)$, 
by simply summing the squared draws from the limiting distributions.
For this illustration we have used $q=3$,
with $2^q=8$ submodels, $Q$ equal to the identity matrix, 
$\tau_0$ equal to 0.1357, and the $\delta$ set to $(0.3,-0.1,1.5)^\tr$. 
The red line represents the scheme where one always 
chooses the widest model. In that case the focus estimator 
is unbiased and its distribution is a perfect normal (as we see). 
The two blue lines are {\it model selection} strategies, 
where a single model is chosen, either using the classic AIC 
(light blue), or using our new median-FIC score (dark blue). 
We see that both strategies induce some bias in the final 
estimator, and that the distribution of $\hatt\mu^*$ 
is a complicated nonlinear mixture of normals. The two green 
lines are {\it model averaging} strategies. The light green 
one is the scheme with weights as in (\ref{eq:lambdaweights}), 
with $\lambda=1$. The dark one is a strategy making use 
of the confidence distributions for the FIC score,
with $\lambda_S$ as in (\ref{eq:coollambda}). 

For this particular position in the $\delta$ parameter space 
the two model averaging strategies produce the most precise 
estimators, obtaining limiting root-mse values of about 1.26 and 1.58
for the average of median-FIC and average with CD-FIC weights. 
The limiting root-mse values for the method selecting 
the best estimator according to the best median-FIC score 
or best AIC scores are respectively 1.60 and 1.67.
The strategy of always selecting the wide model 
has a limiting rmse of 1.74, 
and is thus the least precise strategy among the five 
for this position in the parameter space. 

\section{Performance aspects for the different versions of FIC}
\label{section:comparison}

Our FIC procedures use estimates of root mean squared errors 
to compare and rank candidate models, and as we have 
demonstrated also lead to informative FIC plots 
and CD-FIC plots. There are several issues and aspects
regarding performance, including these: 
(a) How good is the root-FIC score, as an estimator
of the rmse? 
(b) How well-working is the implied FIC scheme for 
finding the underlying best model,
e.g.~as a function of increasing sample size?   
(c) How precise is the final estimator, which would
be the after-selection estimator 
$\hatt\mu_\final$ of (\ref{eq:averaging}) 
or more generally the model average estimator $\hatt\mu^*$ 
of (\ref{eq:mastermaster})? We note that themes (b) and (c)
are quite related, even though different specialised
questions might be posed and worked with to address
particularities. Also, in various contexts, theme (c) is 
`the proof of the pudding' issue. 

Methods to be compared are the unbiased $\fic^u$,
the truncated $\fic^t$, the median-FIC $\fic^m$,
and also its more general variant the quantile-FIC $\fic^q$. 
Themes (a), (b), (c) can of course be studied 
for finite sample sizes, in different setups and with 
many variations. It is again illuminating and indeed 
simplest to travel to the limit experiment setup of 
Sections \ref{section:basics}-\ref{section:CDforfic}, 
however, where complexities are stripped down to the basics, 
with certain basis parameters given and the crucial 
relative distance parameter $\delta=\rootn(\gamma-\gamma_0)$
estimated via a single $D\sim\N_q(\delta,Q)$.  
Below we report on relatively brief investigations
into themes (a), (b), (c). 

\subsection{FIC for estimating mse.} 

\begin{figure}[h]
\centering
\includegraphics[scale=0.45]{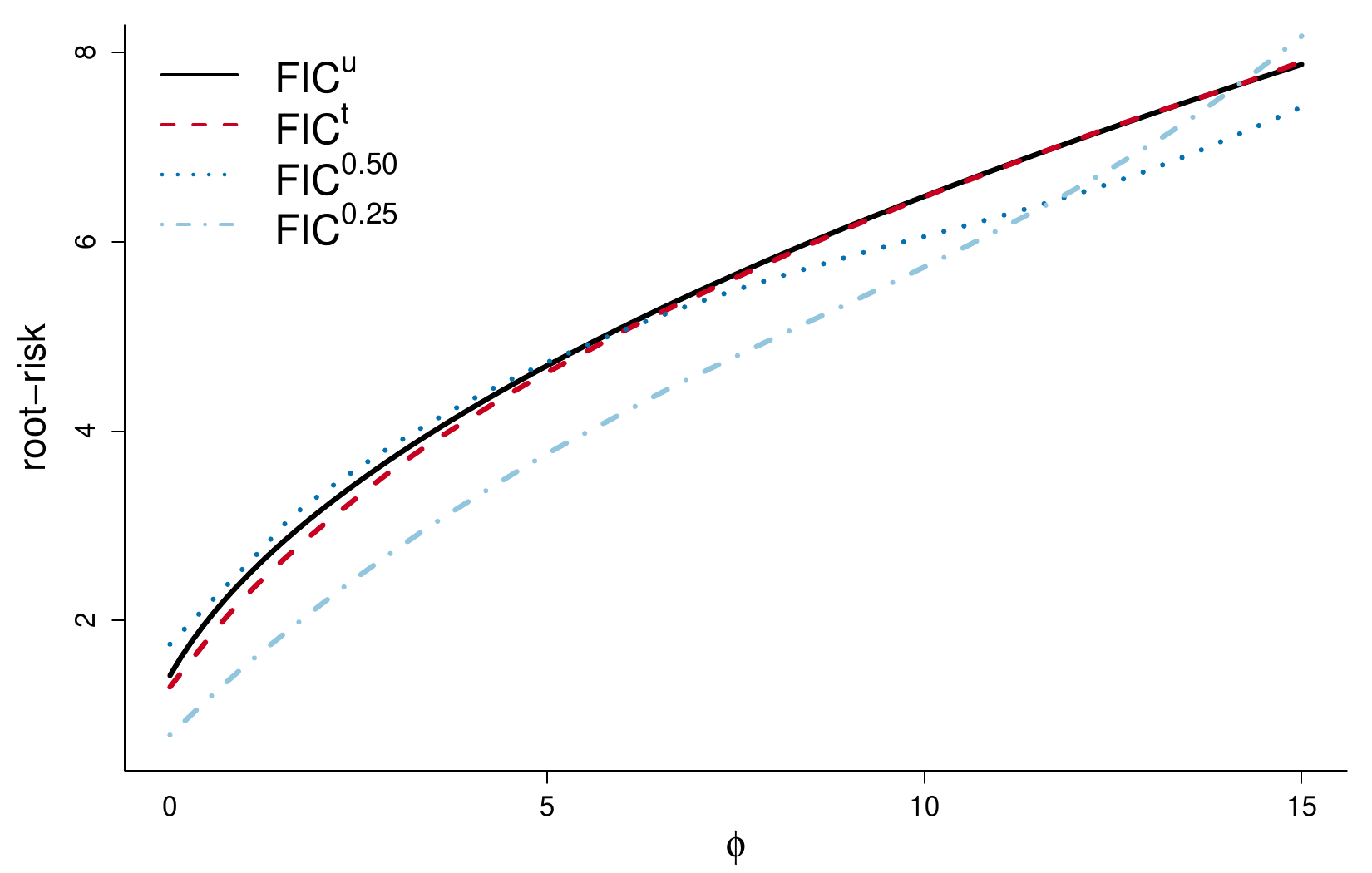}
\caption{Root-mse risk functions $\risk(\phi)^{1/2}$,
for four estimators of $\phi=\eta^2$ is the setup 
where $X\sim\N(\eta,1)$. These correspond to 
the unbiased $\fic^u$, the truncated $\fic^t$, 
and two version of the quantile-FIC $\fic^q$, 
with $q=0.50$ and $q=0.25$.}
\label{figure:fig18}
\end{figure}

The limiting $\mse$ expressions are of the form 
$\tau_S^2+(a_S\delta)^2$, say, as per (\ref{eq:mseS}), 
with $\tau_S$ and $a_S$ known quantities. 
The different FIC schemes differ with respect to 
how the squared bias term is estimated.
In the reduced prototype form worked with at 
the start of Section \ref{section:CDforfic}, 
the comparison boils down to investigating 
four methods for estimating $\phi=\eta^2$ 
in the setup with a single $X\sim\N(\eta,1)$. 
The unbiased and truncated FIC are associated with 
the estimation schemes $\hatt\phi^u=X^2-1$
and $\hatt\phi^t=\max(X^2-1,0)$, whereas the 
median-FIC corresponds to setting $\hatt\phi^m$
equal to the median of the confidence distribution 
$C(\phi,x)=1-\Gamma_1(x^2,\phi)$. Risk functions 
$\risk(\phi)=\E_\phi\,(\hatt\phi-\phi)^2$ can 
now be numerically computed and compared, for the 
different estimators, yielding say 
$\risk^u(\phi),\risk^t(\phi),\risk^m(\phi),\risk^q(\phi)$; 
the first is incidentally equal to $2+4\phi$. 
Figure \ref{figure:fig18} displays four root-risk
functions, i.e.~$\risk(\phi)^{1/2}$. We learn that the 
two `usual' FIC based methods, the unbiased and 
truncated, are rather similar, though the truncated
version is uniformly better for this particular task. 
The quartile-FIC is significantly better for 
a relatively large window of squared bias values, 
whereas the median-FIC is better when such values
are large. 

\subsection{Narrow vs.~wide.}
\label{subsection:narrowwide} 

We now consider a relatively simple set-up, where we 
only wish to choose between two models, 
the narrow (with $p$ parameters) and the wide 
(with $p+q$ parameters). The limiting mean squared errors are 
\beqn
\mse_\narr=\tau_0^2+(\omega^\tr\delta)^2 
   \quadandquad
\mse_\wide=\tau_0^2+\omega^\tr Q\omega, 
\eeqn
from which it also follows that the narrow model 
is better than the wide in the infinite band 
$|\omega^\tr\delta|\le (\omega^\tr Q\omega)^{1/2}$. 
The FIC in effect attempts to use data to see whether 
$\delta$ is inside this band or not. We have 
\beqn
\fic_\narr^u&=&\tau_0^2+(\omega^\tr D)^2-\omega^\tr Q\omega, \\
\fic_\narr^t&=&\tau_0^2+\max\{(\omega^\tr D)^2-\omega^\tr Q\omega,0\}.
\eeqn 
Thus the unbiased $\fic^u$ says that the narrow is best 
if and only if $|\omega^\tr D|\le\sqrt{2}(\omega^\tr Q\omega)^{1/2}$,
and a bit of analysis reveals that the truncated 
$\fic^t$ in this case is in full agreement. 
In the limit experiment of this two-models setup, 
$\psi=\omega^\tr\delta$ has the estimators 
$\hatt\psi_\narr=0$ and $\hatt\psi_\wide=\omega^\tr D$, 
and the final estimator used is 
\beqn
\hatt\psi_\final=
\begin{cases}
\hatt\psi_\narr &{\rm if\ } |t(D)|\le\sqrt{2}, \\
\hatt\psi_\wide &{\rm if\ } |t(D)|>\sqrt{2}. 
\end{cases}
% \quad {\rm where\ }
% t(D)={\omega^\tr D\over (\omega^\tr Q\omega)^{1/2}}. 
\eeqn 
Here 
\beq
\label{eq:tDandeta}
t(D)={\omega^\tr D\over (\omega^\tr Q\omega)^{1/2}}, 
\quad {\rm which\ has\ distribution\ }\N(\eta,1)
{\rm\ with\ }\eta={\omega^\tr\delta\over (\omega^\tr Q\omega)^{1/2}}. 
\eeq 
This FIC strategy is then to be contrasted with 
that of the median-FIC. The question is when 
\beqn
\fic_\narr^m
%% =C_\narr^{-1}(\half)
   =\min\{\mse_\narr\colon C_\narr(\mse_\narr)\ge\half\}
   \le \tau_0^2+\omega^\tr Q\omega, 
\eeqn 
where 
\beqn
C_\narr(\mse_\narr)=1-\Gamma_1\Bigl({(\omega^\tr D)^2\over \omega^\tr Q\omega},
   {\mse_\narr-\tau_0^2\over \omega^\tr Q\omega}\Bigr) 
   \quad {\rm for\ }\mse_\narr\ge\tau_0^2. 
\eeqn 
This means finding when the function $1-\Gamma_1(t(D)^2,1)$ crosses 0.50,
and a simple investigation shows that $\fic^m$ prefers 
the narrow to the wide model if and only if $|t(D)|\le1.0505$. 

The limiting risk functions for the three FIC methods
of reaching a final estimator $\hatt\mu_\final$ are therefore
of the form 
\beqn
\risk(\delta)=\tau_0^2+\omega^\tr Q\omega\,R(\eta),
   \quad {\rm with} \quad
R(\eta)=\E_\eta\,[ I\{|t(D)|>t_0\}t(D)-\eta]^2, 
\eeqn 
using (\ref{eq:tDandeta}), with cut-off value 
$t_0=\sqrt{2}$ for $\fic^u$ and $\fic^t$, 
and with $t_0=1.0505$ for $\fic^m$. 
More generally, the quantile-FIC method of (\ref{eq:quantilefic}) 
can be seen to have such a cut-off value 
$t_0=\Gamma_1^{-1}(1-q,1)^{1/2}$, which is e.g.~$t_0=1.6859$
for $q=0.25$.  
The conservative strategy, choosing the wide model
regardless of the observed $D$, corresponds to cut-off value $t_0=0$. 

% G <- function(x)
% {1-pchisq(x^2,1,1)}
% curve(G,0,2) 

Let us also briefly point to the classic AIC method, in this setup.
As shown in \citet[Chs.~5, 6]{ClaeskensHjort08}, 
in the limit AIC prefers the narrow over the wide model
if and only if $D^\tr Q^{-1}D\le 2q$. With notation as 
in Sections \ref{section:basics}--\ref{section:CDforfic}, 
the limit distribution of the AIC selected estimator becomes 
\beqn
\rootn(\hatt\mu_\aic-\mu_\true)\arr_d
   \Lambda_0+(\omega^\tr Q\omega)^{1/2}
   \bigl[\eta-I\{D^\tr Q^{-1}D>2q\}t(D)\bigr]. 
\eeqn 
When there is only $q=1$ extra parameter in the wide model,
this is the very same as for the two first FIC methods,
with cut-off value $t_0=\sqrt{2}$.

\begin{figure}[h]
\centering
\includegraphics[scale=0.45]{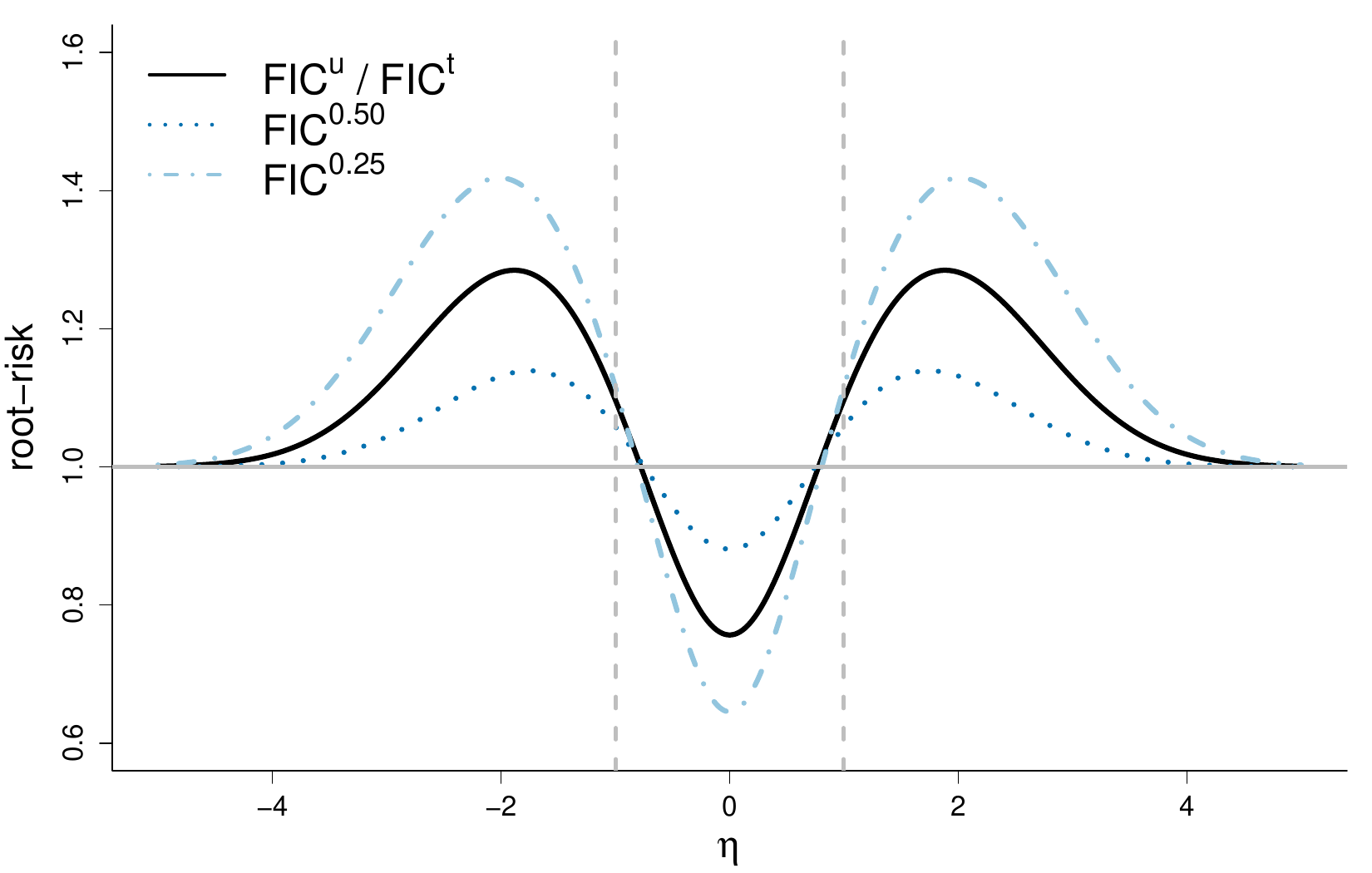}
\caption{For the one-dimensional case $q=1$, 
root-risk function curves for estimators 
coming from three different FIC selection schemes,
as functions of  $\eta=\omega^\tr\delta/(\omega^\tr Q\omega)^{1/2}$:  
the usual FIC (black full),
here also equivalent to the AIC;  
the median-FIC (dotted, blue, and with lowest maximum);
and the quantile-FIC with $q=0.25$ (dotdashed, blue, and with highest maximum). 
Also shown is the benchmark wide procedure (grey, constant).
Inside the two vertical grey lines the narrow model 
is truly better than the wide.}
\label{figure:fig12}
\end{figure}

%% new: 

Figure \ref{figure:fig12} displays root-risk
functions $R(\eta)^{1/2}$
for the usual FIC (with $t_0=\sqrt{2}$, full curve),
for the median-FIC (with $t_0=1.0505$, dotted curve, low max value),
and the quantile-FIC with $q=0.25$ (with $t_0=1.6959$,
dotdashed curve, high max value).
We see that the median-FIC often wins over the standard FIC,
and its maximum risk is considerably lower.
More precisely, median-FIC has the lowest risk in the parts of the
parameter space where the wide model is truly the best model,
but where $\eta$ only has moderately
large values, i.e.~the parts of the parameter space
where the true model is at some moderate distance
from the narrow model.
This fits well with some of our insights from 
Section \ref{section:medianfic}, 
where we saw that median-FIC will select the wide model
with a higher probability than ordinary FIC.
For moderate $\eta$ values, the propensity 
of the median-FIC to select the wide model turns out 
to be good in terms of risk; for $\eta$ values 
farther away from zero all strategies always select 
the wide model and they therefore have identical risk.

For $\eta$ values closer to zero, in the part of the parameter
space where the narrow model is truly more precise than the wide,
we see that median-FIC has a higher risk than the other 
strategies and that quantile-FIC with $q=0.25$ is the 
best strategy. Again this is related to our comments 
in Section \ref{section:medianfic}, 
with $q=0.25$ quantile FIC tending to give lower
FIC scores to the non-wide models, compared to the other strategies.
In this scenario, this gives $\fic^{0.25}$ a propensity 
to select the narrow model. This property is advantageous 
for $\eta$ values around zero, but gives $\fic^{0.25}$ 
a higher risk for moderately large $\eta$ values.

% [xx nils: should round this off suitably. perhaps put in
% in proper place a reminder that these limiting risks
% come from $\rootn(\hatt\mu_\final-\mu_\true)$, with 
% FIC formulae from (\ref{eq:ficficdata}) etc. xx]

\subsection{Three FIC schemes with q = 2.} %% $q=2$
\label{subsection:threeFIC} 

We continue with the somewhat more complex case 
where we have $q=2$ extra parameters in the wide model, 
and four submodels under consideration, here denoted by
0, 1, 2, 12. We let $Q=\diag(\kappa_1^2,\kappa_2^2)$ 
be diagonal, in order to have simpler expressions 
than otherwise. 
This in particular means that $\hatt\gamma_1$ and $\hatt\gamma_2$ 
become independent in the limit.  
The $\mse$ expressions for the four different candidate 
models are then 
\beqn
\mse_0&=&\tau_0^2+(\omega_1\delta_1+\omega_2\delta_2)^2, \\
\mse_1&=&\tau_0^2+\omega_1^2\kappa_1^2+\omega_2^2\delta_2^2, \\
\mse_2&=&\tau_0^2+\omega_2^2\kappa_2^2+\omega_1^2\delta_1^2, \\
\mse_{12}&=&\tau_0^2+\omega_1^2\kappa_1^2+\omega_2^2\kappa_2^2,
\eeqn 
where $\tau_0,\omega_1,\omega_2,\kappa_1,\kappa_2$ are 
considered known parameters, whereas what one can know
about $\delta=(\delta_1,\delta_2)$ is limited to 
the independent observations
$D_1\sim\N(\delta_1,\kappa_1^2)$ and 
$D_2\sim\N(\delta_2,\kappa_2^2)$. 
The FIC scores $\fic^u$, $\fic^t$, $\fic^m$ 
will depend on these known parameters and on $D=(D_1,D_2)^\tr$, 
and the associated limiting risks will be 
functions of $\delta=(\delta_1,\delta_2)$,
\beqn
\risk(\delta)
 =\E_\delta\,|\Lambda_0+\omega^\tr\{\delta-\hatt\delta(D)\}|^2 
 =\tau_0^2 + \E_\delta\,\{\omega^\tr\hatt\delta(D)-\omega^\tr\delta\}^2 
\eeqn
for the three different versions of 
\beqn
\hatt\delta(D)=v_0(D)\begin{pmatrix} 0 \\ 0 \end{pmatrix}
               +v_1(D)\begin{pmatrix} D_1 \\ 0 \end{pmatrix}
               +v_2(D)\begin{pmatrix} 0 \\ D_2 \end{pmatrix}
               +v_{12}(D)\begin{pmatrix} D_1 \\ D_2 \end{pmatrix}, 
% =
%\begin{pmatrix}
% \{v_1(D)+v_{12}(D)\}D_1 \\
% \{v_2(D)+v_{12}(D)\}D_2
% \end{pmatrix}, 
\eeqn 
with $v_0(D),v_1(D),v_2(D),v_{12}(D)$ the associated 
indicator functions for where submodels 0, 1, 2, 12 
are selected. 

\begin{figure}[h]
\centering
\includegraphics[scale=0.3]{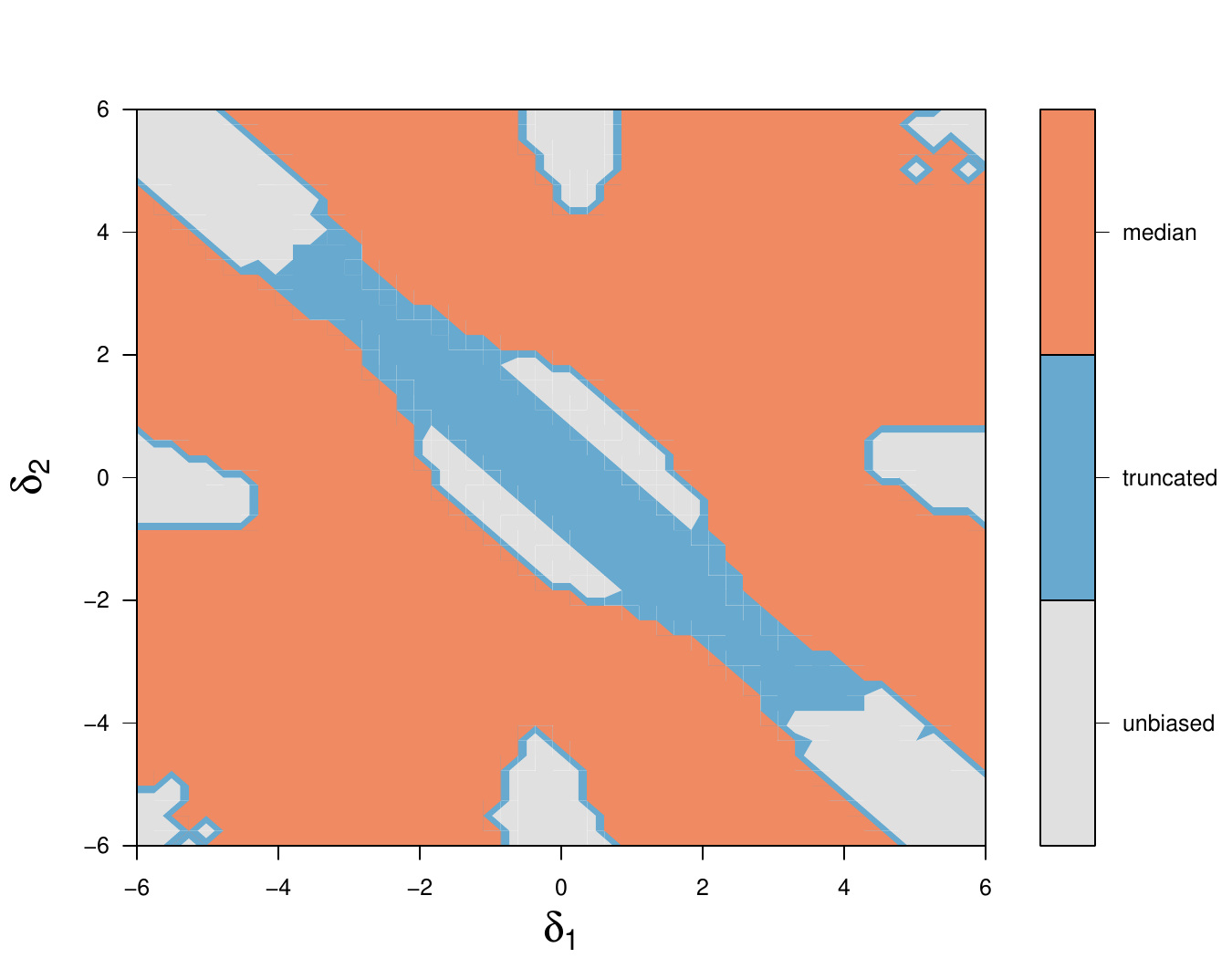}
\includegraphics[scale=0.3]{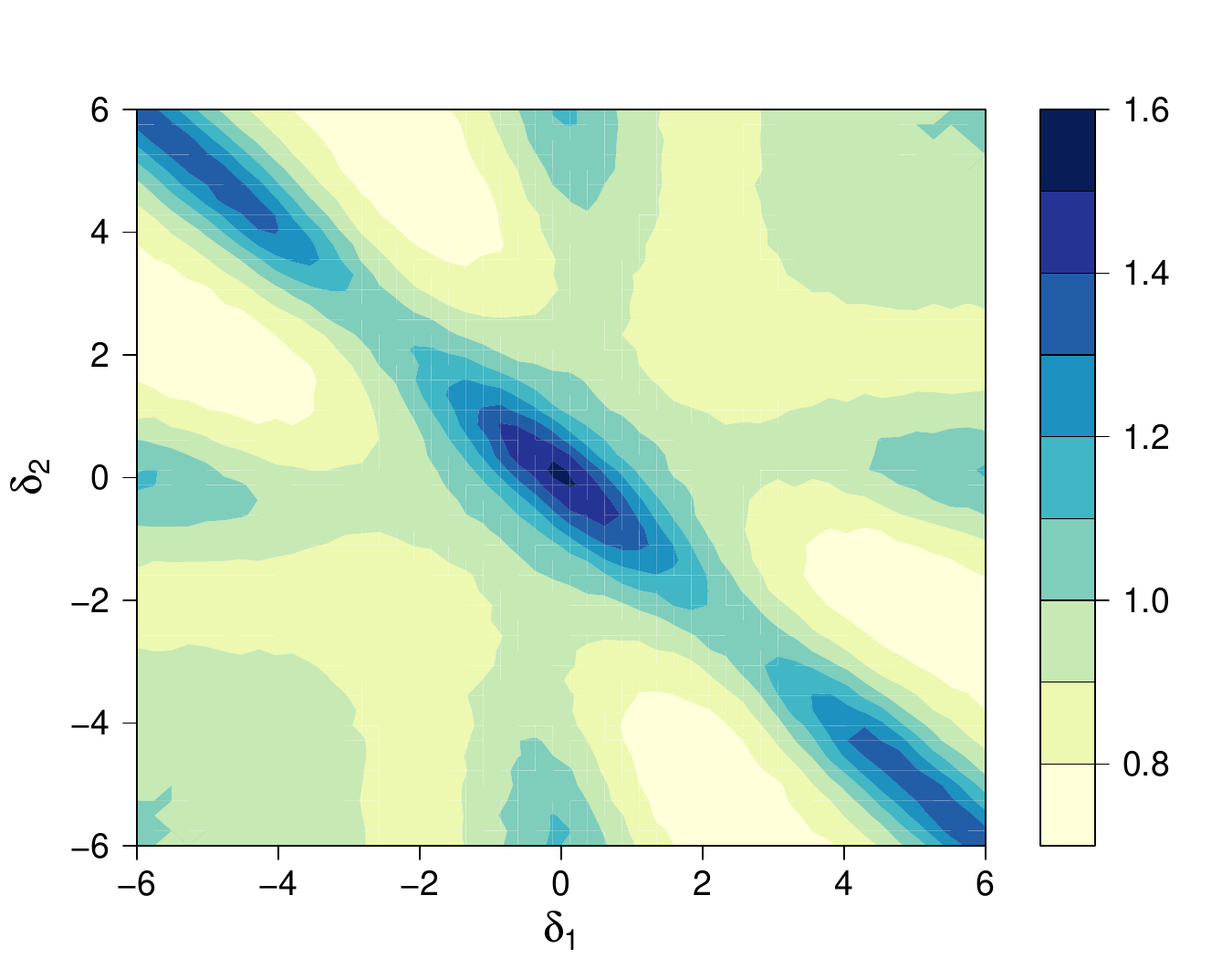}
\caption{Contour plots for the risk depending on the 
true value of $\delta_1$ and $\delta_2$. Left: showing 
which of the three FIC scores gives the estimator with 
the smallest risk; median-FIC is the winner in the red 
area. Right: the median-FIC risk divided by the minimum 
risk among the two competing strategies.}
\label{figure:fig6}
\end{figure}

We can now compute and compare these risk functions in the 
two-dimensional $\delta$ space, for each choice of 
$\tau_0,\omega_1,\omega_2,\kappa_1,\kappa_2$. 
Since the mse expressions, as well as the risk functions, 
all have the same $\tau_0^2$ term, we disregard that contribution,
and in effect set $\tau_0=0$. 
In Figure~\ref{figure:fig6} we show the results of such 
+an exercise, with $\omega=(1,1)^\tr$ and $\kappa=(1,1)^\tr$. 
On the left hand side, we see that for this set-up median-FIC 
gives lower risk than the two other strategies for a relatively 
large part of the parameter space. The right side shows 
the ratio between the risk of median-FIC and the best 
competing strategy. 
The panels indicate that median-FIC beats the two 
other strategies for moderate values of both 
$\delta_1$ and $\delta_2$, but loses when one or both of 
these quantities are close to zero, and also when both 
are large in absolute size. 

This is consistent with our observations in 
Section \ref{section:medianfic}; the median-FIC 
has good performance in the parts of the parameter space 
where the wide model is truly the best.
If quantile-FIC with $q=0.25$ had been included in this comparison,
we would have discovered that $\fic^{0.25}$ beats the 
other strategies in the areas were the wide model is not the best, 
particularly in the narrow diagonal band from $(-6,6)$ to $(6,-6)$.
%% celine hvilket band er dette 

\begin{figure}[h]
\centering
\includegraphics[scale=0.5]{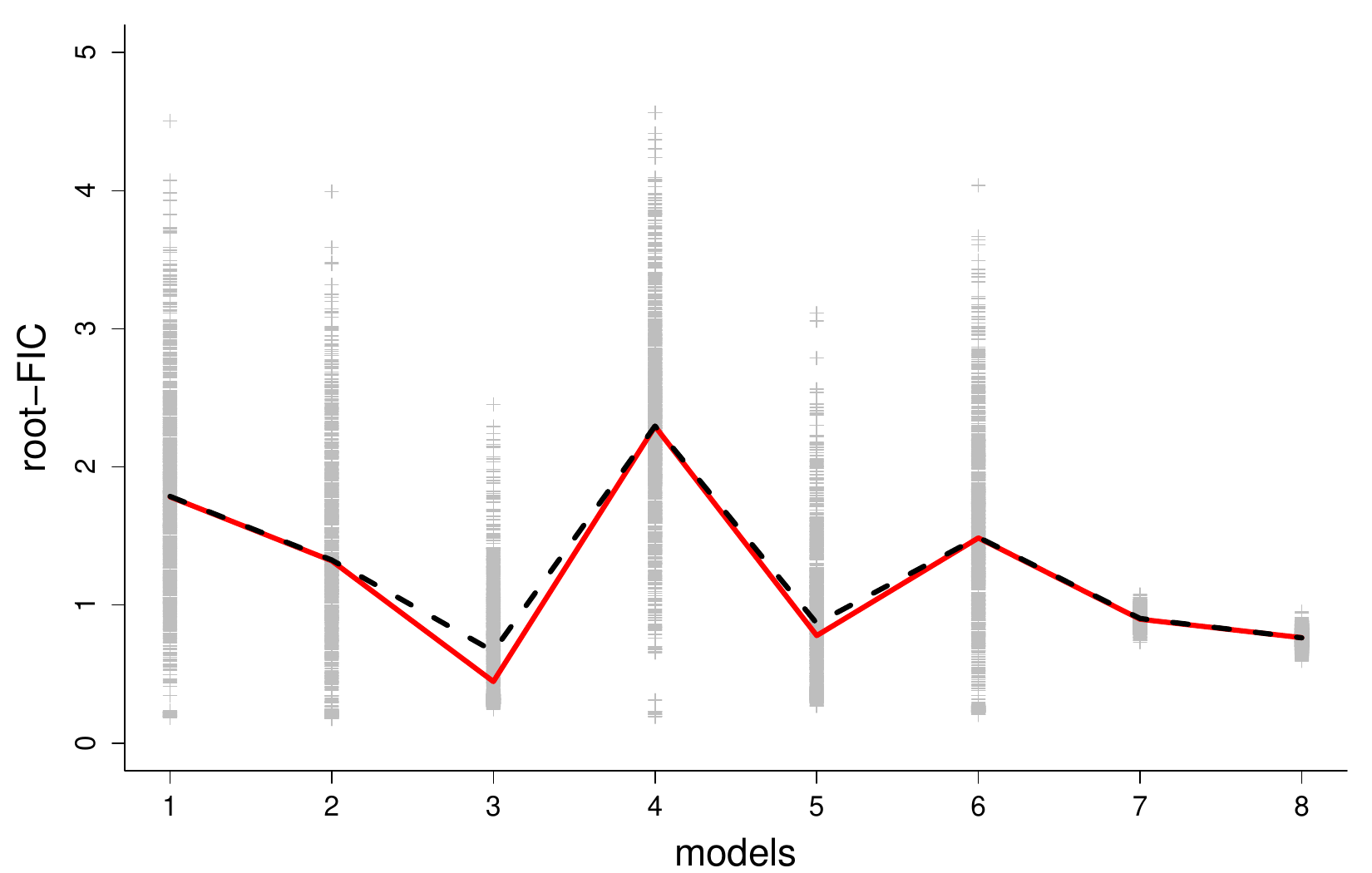}
\caption{Simulation results for finite-sample performance 
in an ordinary linear model. The red line indicates 
the true rmse values, the grey crosses are the root-FIC 
scores from $10^3$ simulated datasets and the black 
dashed lines are the average root-FIC scores.}
\label{figure:fig24}
\end{figure}

\subsection{Finite-sample performance evaluations.} 
\label{subsection:finitesample} 

We have also conducted various investigations of the performance 
of the FIC scores and of the CDs in finite-sample settings. 
In these experiments we sample data from a known wide model 
and with a particular choice of focus parameter, for which 
we then know the true value. 
% We first conduct one round of simulations where we generate 
% a large number of datasets (here $10^4$), 
% let all the candidate models estimate the focus, 
% and then use these estimates to compute the true underlying 
% rmse for each model 
% (or, more precisely, a very good estimate of the true rmse).
Our illustration here is for the linear normal model,
say $y_i=x_i^\tr\beta+z_i^\tr\gamma+\eps_i$ 
with errors being independent from the $\N(0,\sigma^2)$, 
and with focus parameter of the type 
$\mu_0=x_0^\tr\beta+z_0^\tr\gamma$. We may then 
work out exact formulae for the root-mse of 
the different candidate model based estimators 
$\hatt\mu_S=x_0^\tr\hatt\beta_S+z_{0,S}\hatt\gamma_S$.  
%% Then, in the second round of simulations we generate new datasets 
We generate a high number of datasets 
from this model, and compute FIC scores and CDs 
for each of these. From this we can investigate aspects 
(a) and (b) mentioned in the beginning of the section. 
Do the root-FIC scores succeed in estimating the true rmse? 
And do the FIC scores provide a correct ranking of the models? 
Further, we can investigate the coverage properties 
of our CDs: do the confidence intervals we obtain from 
the CD, say the 80\% intervals, cover the true rmse values 
for approximately 80\% of the rounds?

In Figure \ref{figure:fig24} 
we display results of such an investigation for 
such a linear normal regression model with 
intercept parameter $\beta$ protected and 
three extra parameters $\gamma_1,\gamma_2,\gamma_3$ 
associated with three covariates considered 
for ex- or inclusion. In the figure $M_1$ is 
the narrow model, with only the intercept, 
and $M_8$ is the wide model with the intercept 
term and all three covariates. 
The other candidate models correspond to including 
or excluding the three covariates. We have used 
$n=100$, $\beta=0$, $\gamma=(2.0,-1.0,0.5)^\tr$, 
and residual standard deviation $\sigma=2$.
The covariates are drawn from a multivariate normal 
distribution with zero means and relatively 
high levels of correlation; 
$\corr(X_1, X_2) = -0.7$, $\corr(X_1, X_3) = -0.7$, 
$\corr(X_2, X_3) = 0.9$. 
The focus parameter is 
$\mu_0 = x_0^\tr\beta+z_0^\tr\gamma$ with 
$x_0=0$ and $\gamma=(-0.1,1.0,-0.5)^\tr$. 
The red line indicates the true rmse values 
for the eight models. The grey crosses are the 
root-median-FIC scores evaluated in $10^3$ 
%% (different) 
datasets. The black dashed line gives 
the average scores from these $10^3$ datasets. 
The realised coverage of the computed 80\% confidence 
intervals are given in Table \ref{table:table24}. 
That table also reports the percentage 
of rounds where each model has the lowest FIC score 
(i.e.~the winning model). In this setup, 
model 3 had the lowest true rmse (as we see in the figure).

\begin{table}[h]
\small
\caption{Simulation results for the linear model. The realised 
coverage of 80\% confidence intervals for mse and the 
percentage of rounds where each model has the lowest FIC score
(i.e.~the winning model). } 
\begin{center}
\begin{tabular}{ccccccccc}  \toprule
& $M_1$ & $M_2$  & $M_3$ &  $M_4$ & $M_5$ & $M_6$  & $M_7$ &  $M_8$ \\ \midrule
    80\% CI & 80.0 & 80.1 & 80.6 & 80.1 &  78.4 & 80.1 & 55.4 & -- \\ 
    winning  \% &  4.1 & 9.8 & 53.9  & 0.1 & 24.3 & 0.5 & 0.0 & 7.3 \\  \bottomrule
\end{tabular}
\label{table:table24}
\end{center}
\end{table}  

\noindent 
From the figure we see that average root-FIC scores 
are close to the true rmse values. Models 3, 5, 8 were 
truly the best models for this focus parameter, and they 
were also selected most of the time (see the table). 
The realised coverage for the 80\% confidence intervals 
for the mse was generally good, except for models 7 and 8. 
The wide model 8 has by construction no uncertainty. 
The CDs for $M_7$ were typically very steep, 
which then leads to overly narrow confidence intervals.  

We have conducted similar investigations for other classes 
of regression models, the logistic and the Poisson, 
again examining the extent to which CDs 
for root-mse quantities work well and whether the resulting
FIC schemes find the best models. In such models
there is no formula for the exact root-mse, but
such values are easily found numerically via simulation.
A more pertinent difference is however the following,
when comparing the linear model with e.g.~Poisson regression.
Our $\mse_S$ expressions (\ref{eq:mseS}), 
used repeatedly in our paper as consequences of the 
$\|\gamma-\gamma_0\|=O(q/\rootn)$ local neighbourhood model 
framework of (\ref{eq:ftrue}), {\it are exact} for linear 
functions of means in the linear model
(see \citet[Section 6.7]{ClaeskensHjort08}), 
but are otherwise to be seen as good approximations
valid when the models are within a reasonable vicinity
of each other. For logistic and Poisson regression
models, therefore, we arrive at tables and figures 
resembling those above, but only if models 
are in the territory of $O(q/\rootn)$ around the narrow model. 
See in this connection Remark C in Section \ref{section:concluding}. 
It is tentatively comforting that the FIC schemes
tend to pick the right models, even outside that
framework, i.e.~even if the root-FIC scores themselves
do not aim at the real root-mse quantities. 

\section{Illustration: Birds on 73 British and Irish Islands}
\label{section:applications}

\citet{Reed81} analysed the abundance of landbirds on 73 
British and Irish islands. In the dataset, characteristics 
of each island were recorded: 
the distance from mainland ($x_1$), 
the log area ($x_2$), 
the number of different habitats ($z_1$), 
an indicator of whether the islands is Irish or British ($z_2$), 
latitude ($z_3$), 
and longitude ($z_4$).
As the notation indicates, we do take $x_1,x_2$ as 
protected covariates, to be included in all candidate models, 
whereas $z_1,z_2,z_3,z_4$ are open. Based on general
ecological theory and study of similar questions 
we also include two potential interaction terms, 
viz.~$z_5=x_2z_1$ and $z_6=x_1x_2$. 
Of the $2^6=64$ candidate models, corresponding to 
inclusion and exclusion of $z_1,\ldots,z_6$, 
we only allow the interaction term $z_5=x_2z_1$ 
in a model if $z_1$ is also inside; this 
leaves us with $64-18=48$ candidate models below.  

Suppose we take an interest in predicting the number 
of species $y_i$ on the Irish island of Cape Clear. 
In Reed's dataset we have the following information 
about this island: it is located at 6.44 km from the mainland, 
at 51.26 degrees north and $-9.37$ degrees east, 
with an area of 639.11 hectares. 
%% and a maximum elevation of 133.4 m. 
At the time of study it had 20 different habitats ($z_1$), 
and 40 different bird species ($y_i$) were observed. 
Assume that we know that the number of habitats has
decreased to 15 -- which model gives the most precise 
estimate of the current number of species? 
(Naturally, all other covariates are unchanged.)

\begin{figure}[h]
\centering
\includegraphics[scale=0.5]{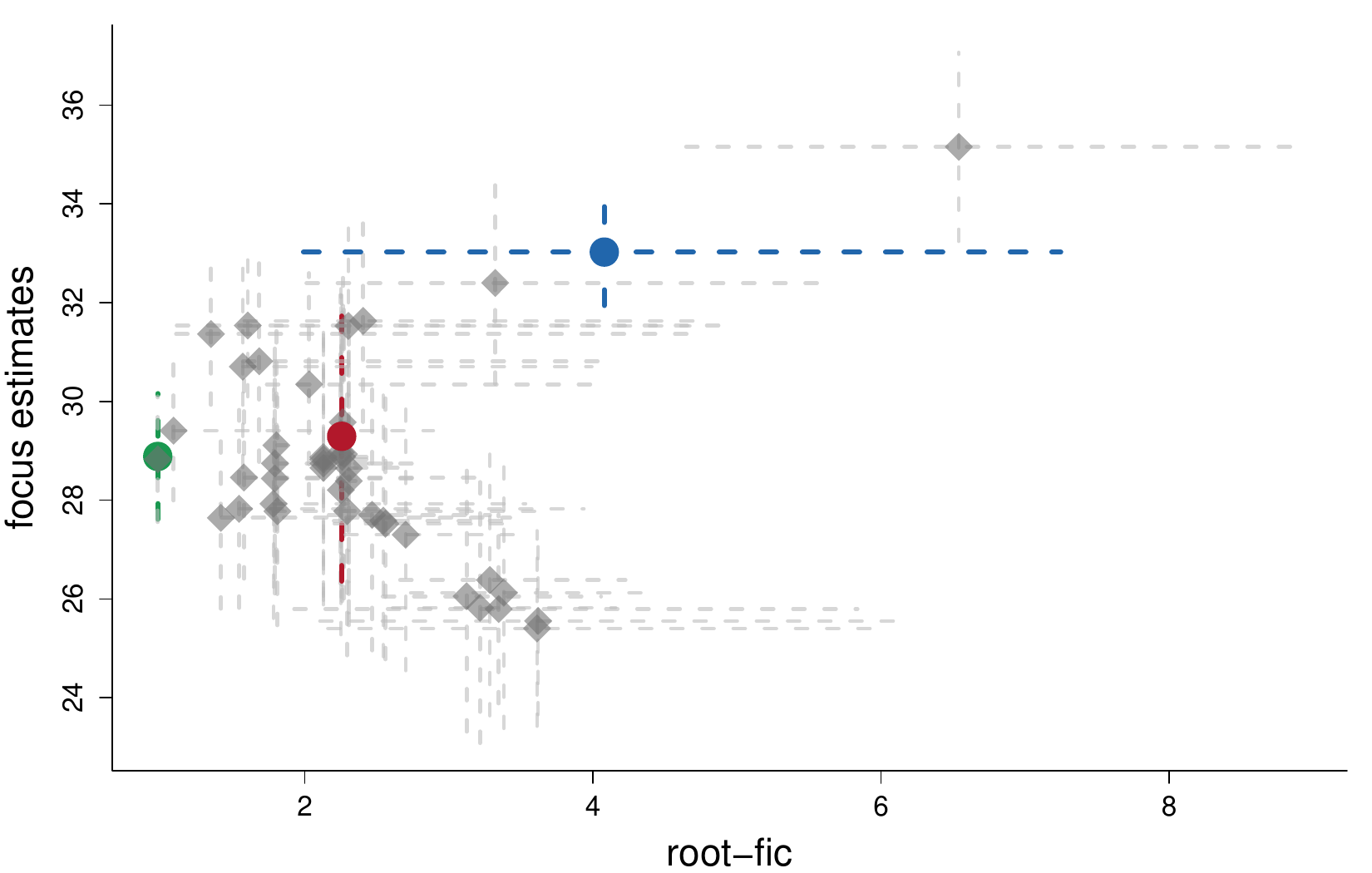}
\caption{FIC plot with associated uncertainty for the 48 
candidate models for estimating the  number of bird species 
on the Irish island Cape Clear. In red the wide model, 
in blue the narrow model, and in green the winning model. 
The uncertainty is represented by 80\% confidence 
intervals. The points plotted are the 
%% `ordinary' 
usual truncated FIC scores, on the $\fic^t/\rootn$ scale.}
\label{figure:fig7}
\end{figure}

As the required wide model we choose the Poisson 
regression model, with 
$y_i\sim\pois(\lambda_i)$, where 
\beqn
\lambda_i 
   = \exp(\beta_0+\beta_1x_{i,1}+\beta_2x_{i,2}
   +\gamma_1z_{i,1}
   +\gamma_2z_{i,2}
   +\gamma_3z_{i,3}
   +\gamma_4z_{i,4} 
   +\gamma_5z_{i,5} 
   +\gamma_6z_{i,6}). 
\eeqn 
%% with the interaction terms $z_1=x_2z_1$ and $z_6=x_1x_2$. 
The wide model thus has nine parameters to estimate, 
while the smallest, narrow one only has three. 
We conduct our FIC analysis, and using our confidence
distribution apparatus we obtain our extended FIC plot 
with uncertainty bands in Figure \ref{figure:fig7}. 
Some models indicate a clear improvement compared 
to the wide model, with very low uncertainty around 
their FIC scores. The winning model is similar to 
the narrow model, but includes the habitat covariate. 
Most of the models with low FIC scores contain 
this covariate, and one or both interaction terms 
or the longitude covariate (Cape Clear lies quite far west 
compared to most of the islands in the dataset).
The predicted number of species on Cape Clear among 
the favoured models is around 29, a decrease from the 
40 species in the dataset. 

% (consistent with the habitat destruction in our hypothetical scenario).

Brief numerical investigations indicate that the coverage properties 
for the confidence intervals for the FIC score are adequate 
when datasets of similar size are generated from the fitted 
wide model above. The uncertainty assessment in 
Figure \ref{figure:fig7} can therefore be trusted 
(even though they are only valid in the limit experiment).

The point to convey with this application is also 
that any other focused statistical question of interest
can be worked with in the same fashion. Natural
focus parameters could be the probability that 
$y$ falls below a threshold $y_0$, given 
a set of present or envisaged island characteristics, 
or the mean function $\E\,(y\midd x_1,x_2,z_1,z_2,z_3,z_4)$
itself, for a given set of covariate combinations. 
For each such focused question, a FIC analysis can be 
run, leading to FIC plots and finessed CD-FIC plots 
as in Figure \ref{figure:fig7}, perhaps each time with 
a new model ranking and a new model winner. 

\section{Discussion}  
\label{section:discussion}

Our paper has extended and finessed the theory of FIC,
through the construction of confidence distributions
associated with each point $(\fic_S^{1/2},\hatt\mu_S)$ 
in the traditional FIC plots and FIC tables. The resulting 
CD-FIC plots enable the statistician to delve deeper 
into how well some candidate models compare to others; 
not only do some parameter estimates have less variance 
than others, but some estimates of the underlying 
root-mse quantities, i.e.~the root-FIC scores, 
are more precise than others. 
The extra programming and computational cost 
is moderate, if one already has computed the usual 
FIC scores. Check in this regard the R package {\tt fic},
which covers classes of traditional regression models; 
see \citet{Jackson19}. 
 
Differences in AIC scores have well-known limiting 
distributions, under certain conditions, 
%% [xx well, it isn't that easy, is it xx] 
which helps users to judge whether the AIC scores 
of two models are sufficiently different as to prefer 
one over the other. Aided by results of our paper 
one may similarly address differences in FIC scores, 
test whether two such scores are significantly
different, etc.; see Remark B in the following section. 

We trust we have demonstrated the usefulness of 
our methodology in our paper, but now point to 
a few issues and perhaps moderate shortcomings. 
Some of these might be addressed in future work; 
see also Section \ref{section:concluding}. 
One concern is that our CDs for root-mse 
are constructed using a local neighbourhood framework
for candidate models, leading to certain mse 
approximations where the squared bias terms 
are put on the same general $O(1/n)$ footing as variances.
First, this is not always a good operating assumption,
and points to the necessity of setting up such 
FIC schemes with care, when it comes to deciding
on the narrow and the wide model, e.g.~which 
covariates should be protected and which open
in the model selection setup. 
Second, the mse approximations, of type (\ref{eq:mseS}),
have led to clear CDs, but where these in essence
stem from accurate analysis of estimated squared biases, 
not fully taking into account the variance parts. 
There is in other words a certain extra layer
of second order variability not brought on board
in this paper's CDs. 
% Since the wide model has no bias, its FIC score 
% has no direct uncertainty, in our setup. 
For any finite dataset, therefore, our CDs will 
to some extent underestimate the true variability 
present in the root-mse estimation.
% The CDs will typically be somewhat too narrow, 
% and the confidence intervals will cover the true rmse 
% with lower probability than indicated [xx ?? xx]. 
Still, we have seen in simulation studies that 
the coverage can be quite reasonable with 
moderate sample sizes, i.e.~that intervals 
of the type $\{\rmse_S\colon C_S(\rmse_S)\le 0.75\}$
have real coverage close to 0.75, etc. 

These issues also mean that the bias associated with 
submodel $S$ will have a strong influence on the 
appearance of the CD for submodel $S$. The CD 
for $\rmse_S$ will start at a position corresponding 
to the estimated variance of that model's
focus parameter estimator $\hatt\mu_S$, 
but the height of the CD at this point will be 
determined by the relative size of the bias, 
viz.~the bias estimate squared divided by the 
variance of the bias. Further, the steepness 
of the CD will mostly be determined by the variance 
of the estimated bias, with a steeper CD 
when the variance of the bias estimate is small.
%% These observations imply the following. 
Thus a particular submodel $S$ will obtain 
a narrow confidence interval around its root-FIC score 
if it leads to a focus estimator with small 
relative bias, or small variance in its bias estimate,
or both. 
% [xx We see this in some examples, but not really so well 
% in Figure 2. The point is that a particular model can 
% get a narrow interval around FIC in two ways: 
% (1) by having a really high point-mass, i.e.~very small 
% relative bias; (2) by having a low point-mass, but 
% a really precise estimate of the bias, i.e.~small $\sigma_S$, 
% then the CD can be really steep.  xx]

% Another potential drawback with our CD construction 
% is that for some models and datasets, the unbiased mse 
% estimate $\fic^u$ will fall outside the 
% confidence curve, and thus outside all confidence 
% intervals too. We see this in Figure \ref{figure:fig21} 
% for the models 000, 001, and 010. 
% [xx not sure what more to say about this xx].

This paper also introduces a new version of the FIC 
score, the quantile-FIC, and its natural special case, 
the median-FIC. One of the benefits of this latter 
FIC score is that it falls directly out of the CD, 
and avoids the need to explicitly decide whether 
one wants to truncate the squared bias or not. 
We have also indicated that the quantile-FIC 
scores can have good performance in large parts 
of the parameter space. More careful examination reveals 
that the advantageous performance of median-FIC 
is primarily found in the parts of the parameter 
space where the wide model really is the most precise. 
These are not the most interesting parameter regions 
when it comes to model selection with FIC, 
because model selection is typically conducted 
in situations where one hopes to find simpler 
effective models than the wide model. 
Our performance investigations reveal that 
other quantile-FIC versions, e.g.~the lower-quartile-FIC 
with $q=0.25$, appears to be a favourable strategy in the 
more crucial parts of the parameter space where the wide model
is outperformed by smaller models.

\section{Concluding remarks} 
\label{section:concluding}

We round off our paper by offering a list of 
concluding remarks, some pointing to further research. 

\smallskip
{\sl A. More accurate finite-sample FIC scores.}
We have extended the FIC apparatus to include 
confidence distributions for the underlying 
root-mse quantities. Our formulae have been developed
via the limit experiment, where there are clear 
and concise expressions both for the mse parameters
and the precision of relevant estimators. For real data
there remain of course differences between 
the actual finite-sample FIC scores, as with 
(\ref{eq:ficficdata}), and the large-sample approximations,
as with (\ref{eq:ficficlimit}). As discussed 
in Section \ref{section:discussion} the CDs 
we construct, based on accurate analysis of limit
distributions, miss part of the real-data variability
for finite samples. It would hence be useful to 
develop relevant finite-sample corrections to 
our CDs. See in this connection also the 
second-order asymptotics section of \citet{HjortClaeskens03b}.

\smallskip
{\sl B. Differences and ratios of FIC scores.}
For two candidate models, say $S$ and $T$ subsets
of $\{1,\ldots,q\}$, our CDs give accurate assessment
of their associated $\rmse_S$ and $\rmse_T$. It would
be practical to have tools for also assessing 
the degree to which these quantities are different. 
It is not easy to construct a simple test for 
the hypothesis that $\rmse_S=\rmse_T$, but 
a conservative confidence approach for addressing 
the mse difference 
\beqn
d(\delta)=\mse_T-\mse_S
   =\tau_T^2-\tau_S^2+\{\omega^\tr(I-G_T)\delta\}^2 
   -\{\omega^\tr(I-G_S)\delta\}^2,  
\eeqn
for any fixed pair of candidate models, is as follows. 
For each confidence level $\alpha$ of interest, 
consider the natural confidence ellipsoid 
$E_\alpha=\{\delta\colon(\delta-D)^\tr Q^{-1}(\delta-D)
\le \Gamma_q^{-1}(\alpha)\}$, with $\Gamma_q^{-1}$
the quantile function for the $\chi^2_q$. 
Then sample a high number of $\delta\in E_q$,
to read off the range $[l_\alpha,u_\alpha]$ or 
values attained by $d(\delta)$. Then the 
confidence of the interval is at least $\alpha$. 
This may in particular be used to construct 
a conservative test for $d(\delta)=0$. 
Similar reasoning applies to other relevant 
quantities, like using ratios of FIC scores to 
build tests and confidence schemes for the 
underlying $\mse_T/\mse_S$ ratios. 

\smallskip
{\sl C. The fixed wide model framework for FIC.}
The setup of our paper has been that of local
neighbourhood models, with these being inside a
common $O(1/\rootn)$ distance of each other. 
This framework, having started with 
\citet{HjortClaeskens03a} and \citet{ClaeskensHjort03},
has been demonstrated to be very useful, 
leading to various FIC procedures in the literature, 
and now also to the extended and finessed FIC
procedures of the present paper. A different and
in some situations more satisfactory framework 
involves starting with a fixed wide model, 
and with no `local asymptotics' involved; 
see the review paper \citet*{ClaeskensCunenHjort19}
for general regression models and 
\citet*{CunenWalloeHjort19} for classes of 
linear mixed models. The key results involve
different approximations to mse quantities, 
along the lines of 
\beqn
\mse_M=\sigma_M^2/n+\{\mu_\true-\mu_M(\theta_{0,M})\}^2,
\eeqn 
for each candidate model $M$. 
Here $\mu_\true$ is defined through the real 
data generating mechanism of the wide model, whereas 
$\mu(\theta_{0,M})$ is the least false parmaeter 
in candidate model, and with $\mu(\theta_M)$ the 
focus parameter expressed in terms of that models's
parameter vector. It would be very useful to lift 
the present paper's methodology to such setups.
This would entail setting up approximate CDs, 
say $C_M(\rmse_M)$, for each candidate model. 
This involves different approximation methods and indeed different 
CD formulae than those worked out in the present paper. 

\smallskip
{\sl D. From FIC to AFIC.}
The FIC machinery is geared towards optimal estimation 
and performance for each given focus parameter.
Sometimes there are several parameters of primary 
interest, however, as with all high quantiles,
or the regression function for a stratum of covariates. 
The FIC apparatus can with certain efforts be lifted
to such cases, where there is a string of focus 
parameters, along with measures of relative importance; 
see \citet[Ch.~6]{ClaeskensHjort08} for such 
average-FIC, or AFIC. The present point is that all
methods of this paper can be lifted to the setting
of such AFIC scores as well. 

\smallskip
{\sl E. Post-selection and post-averaging issues.}
The distribution of post-selection and post-averaging
estimators are complicated, as seen in 
Section \ref{section:modelaveraging}, with limits 
being nonlinear mixtures of normals. Supplementing 
such estimators with accurate confidence analysis
is a challenging affair, 
see e.g.~\citet{Efron14, Hjort14, Kabaila19}. 
Partial solutions are considered in 
\citet[Ch.~7]{ClaeskensHjort08}, \citet{Fletcher19}. 

% \citet{Kabaila19} point to difficulties with constructing
% confidence intervals after averaging; methods 
% able to also take additional variability associated
% with estimated weights are discussed in 
% \citet[Ch.~7]{ClaeskensHjort08} and \citet{Fletcher19}. 

% \smallskip
% {\sl E. Confidence distributions for AIC scores.}
% [xx AIC things. full CD for the population quantity 
% $\delta^\tr Q^{-1}\delta-2q$. we can actually
% make something like a median-AIC. xx]

%%%%%%%%%%%%

\section*{Acknowledgments} 

The authors are grateful for partial support from
the Norwegian Research Council for the five-year 
research group FocuStat (Focused Statistical Inference
with Complex Data, led by Hjort), and they have
benefitted from many FIC and CD long-term discussions
inside that group. 

%% Mention the {\tt fic} package of Jackson's. 

\bibliographystyle{biometrika}
\bibliography{fic_bibliography2020}

\end{document}